\documentclass[12pt]{article}

\usepackage[T2A]{fontenc}
\usepackage[cp866]{inputenc}
\usepackage[russian]{babel}
\usepackage{graphicx}

\usepackage{amsmath}
\usepackage{a4wide}
\usepackage{amsmath,amssymb}

\begin{document}

\noindent {\sl Problems of Information Transmission},\\
\noindent vol. 43, no. 2, pp. 3-24, 2007.

\vskip 0.8cm

\begin{center} {\bf M. V. Burnashev}
%\footnote[1]{ Institute for Information Transmission Problems,
%Russian Academy of Sciences, 19 Bolshoi Karetni,
%101447 Moscow, Russia }
\end{center}

\vskip 0.4cm

\begin{center}
{\large\bf CODE SPECTRUM AND RELIABILITY FUNCTION: GAUSSIAN
CHANNEL \footnote[1]{The research described in this
publication was made possible in part by the Russian Fund for
Fundamental Research (project number 06-01-00226).}}
\end{center}

{\begin{quotation} \normalsize A new approach for upper bounding
the channel reliability function using the code spectrum is
described. It allows to treat both low and high rate cases in a
unified way. In particular, the earlier known upper bounds are
improved, and a new derivation of the sphere-packing bound is
presented.
\end{quotation}}

\vskip 0.7cm

\begin{center}
{\large\bf \S\;1. Introduction and main results}
\end{center}

We consider the discrete time channel with independent additive
Gaussian noise, i.e. if $\mbox{\boldmath $x$} = (x_1,\ldots,x_n)$ is
the input codeword then the received block
$\mbox{\boldmath $y$} = (y_1,\ldots,y_n)$ is
$$
y_i = x_i + \xi_i\,, \quad i=1,\ldots,n \,,
$$
where $(\xi_1,\ldots,\xi_n)$ are independent Gaussian r.v.'s with
${\mathbf E} \xi_i=0\,,\; {\mathbf E} \xi_i^2=1$.

For $\mbox{\boldmath $x$},\mbox{\boldmath $y$} \in
{\mathbb{R}}^{n}$ denote $(\mbox{\boldmath $x$},\mbox{\boldmath
$y$}) = \sum\limits_{i=1}^n x_i y_i,\;\|\mbox{\boldmath $x$}\|^2 =
(\mbox{\boldmath $x$},\mbox{\boldmath $x$}),\,
d\left(\mbox{\boldmath $x$},\mbox{\boldmath $y$}\right) =
\|\mbox{\boldmath $x$} - \mbox{\boldmath $y$}\|^{2}$ and
$S^{n-1}(b) = \{\mbox{\boldmath $x$} \in R^{n}: \|\mbox{\boldmath
$x$}\| = b\}$. We assume that all codewords $\mbox{\boldmath $x$}$
satisfy the condition $\|\mbox{\boldmath $x$}\|^2 = A n$, where
$A>0$ is a given constant. A subset ${\cal C} = \{\mbox{\boldmath
$x$}_{1},\ldots, \mbox{\boldmath $x$}_{M}\} \subset
S^{n-1}(\sqrt{An}),\,\\M = e^{Rn}$, is called a $(R,A,n)$-code of
rate $R$ and length $n$. The {\it minimum distance} of the code
${\cal C}$ is $d({\cal C}) = \min \{d(\mbox{\boldmath $x$}_{i},
\mbox{\boldmath $x$}_{j}):i \neq j\}$.

The channel reliability function \cite{Sh,SGB1} is defined as
$$
E(R,A) = \limsup_{n \to \infty}\,\frac{1}{n}\,
\ln \frac{1}{P_{\rm e}(R,A,n)} \;,
$$
where $P_{\rm e}(R,A,n)$ is the minimal possible decoding error
probability for a $(R,A,n)$-code.

After the fundamental results of the paper \cite{Sh}, further
improvements of various bounds for $E(R,A)$ have been obtained in
[2--9]. In particular, on the exact form of the function $E(R,A)$
it was known only that \cite{Sh}
\begin{equation}\label{exact}
E(0,A) = \frac{A}{4}\,, \qquad
E(R,A)= E_{\rm sp}(R,A)\,, \quad R_{\rm crit}(A) \leq R \leq C(A)\,,
\end{equation}
where
\begin{equation} \label{sphere2}
\begin{gathered}
C=C(A)= \frac{1}{2}\,\ln (1+A)\,, \quad
R_{\rm crit}(A) = \frac{1}{2} \ln \frac{2 + A + \sqrt{A^2+4}}{4} \,,
\end{gathered}
\end{equation}
\begin{equation} \label{sphere1}
\begin{gathered}
E_{\rm sp}(R,A) = \frac{A}{2} -
\frac{\sqrt{A(1-e^{-2R})} g(R,A)}{2} - \ln g(R,A) + R \,, \\
g(R,A) = \frac{1}{2}\left(\sqrt{A(1-e^{-2R})} +
\sqrt{A(1-e^{-2R})+4} \right) \,.
\end{gathered}
\end{equation}

Moreover, recently \cite{BM2} the exact form of $E(R,A)$ for a new
region $\overline{R}_{1}(A) \leq R \leq R_{\rm crit}(A)$ was claimed
under some restriction on $A$. Similar to the case of the binary
symmetric channel (BSC), that assertion follows from a useful
observation that the tangent (it has the slope $(-1)$) to the
function $E_{\rm sp}(R,A)$ at the point $R=R_{\rm crit}(A)$ touches
the previously known upper bound for $E(R,A)$ [5--7]. Since those
results from [5--7] were proved under some restrictions on $A$,
those restrictions were remaining in \cite{BM2} as well. Since there
are some inaccuracies in the formulation of that result in
\cite{BM2} we do not expose corresponding formulas from \cite{BM2}
(moreover, they have a different from ours form).

From theorem 1 and the formula (\ref{exact1}) (see below)
the exact form of $E(R,A)$ follows  for the region
$\overline{R}_{1}(A) \leq R \leq R_{\rm crit}(A)$ for any $A > 0$.
Moreover, if $A > A_{0} \approx 2.288$ (see (\ref{defR2})) then from
theorem 2 below the exact form of $E(R,A)$ follows for a wider
region $\overline{R}_{3}(A) \leq R \leq R_{\rm crit}(A)$, where
$\overline{R}_{3}(A) < \overline{R}_{1}(A)$ and
$\overline{R}_{3}(A) \approx R_{\rm crit}(A) - 0.06866,\,
A \geq A_{0}$.

For $0 < R < \overline{R}_{1}(A),\,0 < A \leq A_{0}$, or
$0 < R < \overline{R}_{3}(A),\,A > A_{0}$, still only lower and
upper bounds for $E(R,A)$ are known [1--9], and in this paper the
most accurate of the upper bounds is improved.

We begin by explaining what constituted the difficulty in upper
bounding the function $E(R,A)$ in the earlier papers [5--9]. Note
that when testing only two codewords
$\mbox{\boldmath $x$}_{i},\mbox{\boldmath $x$}_{j}$ with large
distance
$\|\mbox{\boldmath $x$}_{i} - \mbox{\boldmath $x$}_{j}\|^{2} = d$
we have the decoding error probability $P_{\rm e} \sim e^{-d/8}$.
Let $B_{\rho n}$ be the average number
of each codeword $\mbox{\boldmath $x$}_{i}$ neighbors on the
approximate distance $2A(1-\rho)n$. It was shown in \cite{ABL1} that
for a $(R,A,n)$-code there exists $\rho$ such that
$B_{\rho n} \gtrsim 2^{b(\rho)n}$, where the function $b(\rho) > 0$
is described below, and $2A(1-\rho)n$ does not exceed the best upper
bound (linear programming) for the minimal code distance
$d({\cal C})$. Therefore, if each codeword
$\mbox{\boldmath $x$}_{i}$ has approximately $B_{\rho n}$ neighbors
on the distance $2A(1-\rho)n$, then it is natural to expect that
$P_{\rm e} \gtrsim B_{\rho n}e^{-A(1-\rho)n/4}$ for large $n$
(and not very small $\rho$), i.e. a variant of an additive
{\it lower} bound for the probability of the union of events holds.

The first variant of such additive bound was obtained in \cite{ABL1}
under rather severe \\ constraints on $R$ and $A$. Those results of
\cite{ABL1} have been strengthened in \cite{Bur4, Bur7}, using the
method of [10--12]. However there were still certain constraints on
$R$ and $A$. It should be noted that the investigation of $E(R,A)$
for the Gaussian channel is similar to the investigation of $E(R,A)$
for the BSC. The difference is only that due to the discrete
structure of a binary alphabet some expressions become simpler. For
the BSC the method of \cite{Bur4} was recently \cite{Bur6,Bur8}
further developed. Although the approach of \cite{Bur6,Bur8} is
still based on \cite{Bur4}, some additional arguments allowed the
approach to be essentially strengthened and simplified.

It should also be noted that until the papers \cite{Bur6,Bur8}, all
papers mentioned made use of various variants of the second order
Bonferroni inequalities.

The main aim of this paper is to prove an additive bound without
any constraints on $R$ or $A$. For that purpose the method of
\cite{Bur6,Bur8} is applied. It is also worth noting that
Bonferroni inequalities are not used. This approach allows us to
treat both low and high rate $R$ cases in a unified way. As an
example, in \S\,2 a new derivation of the sphere-packing bound is
presented.

Introduce some notations.  For a code
${\cal C} = \{\mbox{\boldmath $x$}_{1},\ldots,
\mbox{\boldmath $x$}_{M}\} \subset S^{n-1}(\sqrt{An})$ denote
\begin{equation} \label{defrho}
\rho_{ij} = \frac{(\mbox{\boldmath $x$}_{i},
\mbox{\boldmath $x$}_{j})}{An} \,, \qquad
d_{ij} = \|\mbox{\boldmath $x$}_{i} -
\mbox{\boldmath $x$}_{j}\|^2 =  2An(1-\rho_{ij}) \,.
\end{equation}
Below it will be convenient to use the parametric representation
of the transmission rate $R = R(t)$ via the monotonic increasing
function
\begin{equation} \label{defRt}
R(t) = (1+t) \ln (1+t) - t\ln t \,, \qquad t \geq 0\,.
\end{equation}
Consequently, for a rate $R \geq 0$ introduce $t_{R} \geq 0$
as the unique root of the equation
\begin{equation} \label{deftR}
R = R(t_{R}) = (1+t_{R}) \ln (1+t_{R}) - t_{R}\ln t_{R} \,.
\end{equation}
Introduce also the functions
\begin{equation} \label{deftau1}
\tau(t) = \frac{2\sqrt{t(1+t)}}{1+2t}\,, \qquad
\tau_{R} = \tau(t_{R})\,.
\end{equation}
We shall need the values
\begin{equation} \label{deft1}
\begin{gathered}
\overline{t}_{1}(A) = \frac{\sqrt{2 + \sqrt{4+A^2}}-2}{4} \,, \qquad
\overline{\tau}_{1}(A)= \tau(\overline{t}_{1}(A)) =
\frac{A}{2+\sqrt{4+A^{2}}}\,, \\
\overline{R}_{1}(A) = R(\overline{t}_{1}(A)) \,,
\end{gathered}
\end{equation}
where the functions $\tau(t),R(t)$ are defined in (\ref{deftau1})
and (\ref{deftR}). Sometimes below  we shall omit the argument $A$ in
$\overline{t}_{1}(A), \overline{\tau}_{1}(A), \overline{R}_{1}(A)$.

One of the main results of the paper is

\medskip

{T h e o r e m\; 1}. {\it For any $A > 0$ the following
%for the reliability
%function $E(R,A)$ the
relations hold:
\begin{equation} \label{exact1}
E(R,A) = \left\{
\begin{array} {l}
E_{\rm sp}(R_{\rm crit},A) + R_{\rm crit}-R\,, \quad
\overline{R}_{1} \leq R \leq R_{\rm crit}\,,  \\
E_{\rm sp}(R,A)\,, \quad R_{\rm crit} \leq R \leq C \,,
\end{array} \right.
\end{equation}
and
\begin{equation} \label{exact1a}
E(R,A) \leq \frac{A(1-\tau_{R})}{4} + \ln (1+2t_{R}) - R \,, \qquad
0 \leq R \leq \overline{R}_{1}\,,
\end{equation}
where $R_{\rm crit}(A),\overline{R}_{1}(A),\tau_{R}$ and $t_{R}$
are defined in} (\ref{sphere2}), (\ref{deft1}), (\ref{deftau1})
{\it and} (\ref{deftR}), {\it respectively}.

\medskip

{\it Remark} 1. We have
$\overline{R}_{1}(A) < R_{\rm crit}(A)\,,\, A > 0$. Moreover,
$\max\limits_{A}\left\{R_{\rm crit}(A)-\overline{R}_{1}(A)\right\}
\approx 0.06866$, and it is attained for $A = A_{0} \approx 2.288$.

{\it Remark} 2. Note that (see the formulas (\ref{exact1}) and
(\ref{exact1a}) for $R = \overline{R}_{1}$)
\begin{equation}\label{check1}
E_{\rm sp}(R_{\rm crit},A) + R_{\rm crit} =
\frac{A(1-\overline{\tau}_{1})}{4} + \ln (1+2\overline{t}_{1})\,.
\end{equation}
Validity of (\ref{check1}) can be checked using the formulas
(\ref{deftR}), (\ref{deftau1}) and the relations
\begin{equation}\label{userel1}
\begin{gathered}
1+2\overline{t}_{1}= \sqrt{\frac{A}{4\overline{\tau}_{1}}}\,,\qquad
R_{\rm crit} = \frac{1}{2}\ln \frac{1}{1-\overline{\tau}_{1}}\,, \\
A\left(1-e^{-2R_{\rm crit}}\right) = A\overline{\tau}_{1} =
\frac{A}{\overline{\tau}_{1}} -4\,, \quad
g(R_{\rm crit}) = \frac{(1+\overline{\tau}_{1})\sqrt{A}}
{2\sqrt{\overline{\tau}_{1}}}\,.
\end{gathered}
\end{equation}

If $A > A_{0} \approx 2.288$ (see (\ref{defR2})) then the upper
bound (\ref{exact1a}) can be slightly improved, and, moreover, the
validity region of the first of formulas (\ref{exact1}) can be
enlarged to $\overline{R}_{3} \leq R \leq R_{\rm crit}$, where
$\overline{R}_{3}(A) < \overline{R}_{1}(A)$ (see (\ref{defR2})).
To explain the possibility of such an improvement consider the
problem of upper bounding the minimal code distance $\delta(R,n)$ of
a spherical code. The best upper bound for $\delta(R,n)$ was
obtained in \cite{KL} using the linear programming bound. It was
also noticed in \cite[p. 20]{KL} that for $R > 0.234$ a better upper
bound for $\delta(R,n)$ is obtained if the linear programming
bound is applied not directly to the original spherical code, but to
its subcode on a spherical cap. That observation was recently used
in \cite{BHL1} when estimating the code spectrum and the function
$E(R,A)$. Using the approach of \cite{Bur4} an upper bound for
$E(R,A)$ was obtained in \cite{BHL1}. But it is rather difficult to
use that upper bound since it is expressed as an optimization
problem over four parameters. In fact, it is possible to get a more
accurate and rather simple bound that constitutes theorem 2 below.

Introduce the function
\begin{equation}\label{defD}
\begin{gathered}
D(t) = \ln\frac{1+t}{t} - \frac{1}{2\sqrt{t(1+t)}} -
\frac{1}{1+2t}\,, \qquad t > 0\,,
\end{gathered}
\end{equation}
and denote $\overline{t}_{2} \approx 0.061176$ the unique root of
the equation $D(t) = 0$. The equivalent equation (with a sign
misprint) appeared earlier in \cite[p. 20]{KL}. Denote also
\begin{equation}\label{defR2}
\begin{gathered}
\overline{R}_{2} = R(\overline{t}_{2}) \approx 0.2339\,, \qquad
\overline{\tau}_{2} = \tau(\overline{t}_{2}) \approx 0.4540\,, \\
\overline{R}_{3}(A)= R_{\rm crit}(A)+ \overline{R}_{2} +
\frac{1}{2}\ln(1-\overline{\tau}_{2}) \approx
R_{\rm crit}(A) - 0.0687\,, \\
A_{0} = \min\left\{A: \overline{R}_{1}(A) \geq
\overline{R}_{2}\right\} \approx 2.288\,.
\end{gathered}
\end{equation}

The next result strengthens theorem 1 when $A > A_{0}$.

\medskip

{T h e o r e m\; 2}. {\it If $A > A_{0} \approx 2.288$ then
the following relations hold:
\begin{equation} \label{exact11}
E(R,A) = \left\{
\begin{array} {l}
E_{\rm sp}(R_{\rm crit},A) + R_{\rm crit}-R\,, \quad
\overline{R}_{3} \leq R \leq R_{\rm crit}\,,  \\
E_{\rm sp}(R,A)\,, \quad R_{\rm crit} \leq R \leq C \,,
\end{array} \right.
\end{equation}
and
\begin{equation} \label{exact1b}
E(R,A) \leq \left\{
\begin{array} {l}
\dfrac{1}{4}A(1-\tau_{R}) + \ln (1+2t_{R}) - R \,, \qquad
0 < R \leq \overline{R}_{2}\,, \\
\dfrac{1}{4}Aae^{-2R} - \dfrac{1}{2}\ln(2 - ae^{-2R}) -
\dfrac{1}{2}\ln a\,, \qquad
\overline{R}_{2} \leq R \leq \overline{R}_{3}(A)\,,
\end{array} \right.
\end{equation}
where
$a = (1-\overline{\tau}_{2})e^{2\overline{R}_{2}} \approx 0.8717$}.

\medskip

For a comparison purpose we present also the best known lower bound
for the function $E(R,A)$ [1;3, Theorem 7.4.4]
\begin{equation}\label{low1}
E(R,A) \geq \left\{ \begin{array}{l}
  A\left(1-\sqrt{1-e^{-2R}}\right)/4\,, \qquad
  0 \leq R \leq R_{\rm low}, \\
  E_{\rm sp}(R_{\rm crit},A) + R_{\rm crit} - R\,, \qquad
  R_{\rm low} \leq R \leq R_{\rm crit}, \\
  E_{\rm sp}(R,A)\,, \qquad R_{\rm crit} \leq R \leq C(A)\,,
\end{array}
\right.
\end{equation}
where
\begin{equation}\label{defR3}
R_{\rm low}(A) = \frac{1}{2} \ln \frac{2+ \sqrt{A^2+4}}{4} \,.
\end{equation}

Combining analytical and numerical methods it can be shown that for
$A > A_{0}$ we have
\begin{equation}\label{ineqR}
R_{\rm low}(A) < \overline{R}_{2} < \overline{R}_{3}(A) <
\overline{R}_{1}(A) < R_{\rm crit}(A)\,.
\end{equation}

On the figure the plots of upper (\ref{exact11}),(\ref{exact1b})
and lower (\ref{low1}) bounds for $E(R,A)$ with $A = 4$ are
presented.

The paper is organized as follows. In \S 2 the main analytical tool
(proposition 1) is presented and, as an example, the sphere-packing
upper bound is derived. In \S 3 proposition 1 and the code
spectrum are combined in propositions 2--3. In \S 4 (using results
of \S 3 and the known bound for the code spectrum - theorem 3)
theorem 1 is proved. In \S 5 theorem 2 is proved. Proofs of some
auxiliary results are presented in Appendix.

%\newpage

\begin{center}
{\large\bf \S\;2. New approach and sphere-packing exponent}
\end{center}

For the conditional output probability distribution density
$p(\mbox{\boldmath $y$}|\mbox{\boldmath $x$})$ of the input
codeword $\mbox{\boldmath $x$}$ the formula holds
$$
\begin{gathered}
\ln p(\mbox{\boldmath $y$}|\mbox{\boldmath $x$}) =  -\frac{1}{2}\,
d(\mbox{\boldmath $y$},\mbox{\boldmath $x$}) -
\frac{n}{2}\ln(2\pi)\,, \qquad
\mbox{\boldmath $x$},\mbox{\boldmath $y$} \in {\mathbb{R}}^{n}
\end{gathered}
$$
(in a similar formula in \cite{Bur4} there is a misprint - the
minus sign is missing). To describe our approach, we fix a small
$\delta = o(1),\,n \to \infty$, and $s > 0$ and for an output
$\mbox{\boldmath $y$}$ define the set:
\begin{equation}\label{defX}
\mbox{\boldmath $X$}_{s}(\mbox{\boldmath $y$}) = \left\{
\mbox{\boldmath $x$}_{i} \in {\cal C} :|d(\mbox{\boldmath $y$},
\mbox{\boldmath $x$}_{i}) - sn| \leq \delta n \right\}\,,
\qquad \mbox{\boldmath $y$} \in {\mathbb{R}}^{n}.
\end{equation}
All codewords $\{\mbox{\boldmath $x$}_{i}\}$ are assumed
equiprobable. For a chosen decoding method denote
$P(e|\mbox{\boldmath $y$},\mbox{\boldmath $x$}_{i})$
the conditional decoding error probability provided that
$\mbox{\boldmath $x$}_{i}$ was transmitted and
$\mbox{\boldmath $y$}$ was received. Denote
$p_{\rm e}(\mbox{\boldmath $y$})$ the probability distribution
density to get the output $\mbox{\boldmath $y$}$ and to make a
decoding error. Then
$$
\begin{gathered}
p_{\rm e}(\mbox{\boldmath $y$}) =  M^{-1}\sum_{i=1}^{M}
p(\mbox{\boldmath $y$}|\mbox{\boldmath $x$}_{i})
P(e|\mbox{\boldmath $y$},\mbox{\boldmath $x$}_{i}) \geq
 M^{-1}\sum_{\mbox{\small\boldmath $x$}_{i} \in
\mbox{\small\boldmath $X$}_{s}(\mbox{\small\boldmath $y$})}
p(\mbox{\boldmath $y$}|\mbox{\boldmath $x$}_{i})
P(e|\mbox{\boldmath $y$},\mbox{\boldmath $x$}_{i}) = \\
=  M^{-1}(2\pi)^{-n/2}\sum_{\mbox{\small\boldmath $x$}_{i} \in
\mbox{\small\boldmath $X$}_{s}(\mbox{\small\boldmath $y$})}
e^{-d(\mbox{\small\boldmath $y$},\mbox{\small\boldmath $x$}_{i})/2}
P(e|\mbox{\boldmath $y$},\mbox{\boldmath $x$}_{i}) \geq \\
\geq M^{-1}(2\pi e^{s+\delta})^{-n/2}
\sum_{\mbox{\small\boldmath $x$}_{i} \in
\mbox{\small\boldmath $X$}_{s}(\mbox{\small\boldmath $y$})}
P(e|\mbox{\boldmath $y$},\mbox{\boldmath $x$}_{i})
\geq M^{-1}(2\pi e^{s+\delta})^{-n/2}\left[
|\mbox{\boldmath $X$}_{s}(\mbox{\boldmath $y$})|-1\right]_{+}\,,
\end{gathered}
$$
where $[z]_{+} = \max\{0,z\}$ and $|A|$ -- the cardinality of the
set $A$ . For the decoding error probability $P_{\rm e}$ we get
$$
P_{\rm e} = \int\limits_{\mbox{\small\boldmath $y$} \in
{\mathbb{R}}^{n}}p_{\rm e}(\mbox{\boldmath $y$})\,
d\mbox{\boldmath $y$} \geq M^{-1}(2\pi e^{s+\delta})^{-n/2}
\int\limits_{\mbox{\small\boldmath $y$}: \left|
\mbox{\small\boldmath $X$}_{s}(\mbox{\small\boldmath $y$})\right|
\geq 2}\left[\left|\mbox{\boldmath $X$}_{s}(\mbox{\boldmath $y$})
\right|-1\right]\,d\mbox{\boldmath $y$}\,.
$$
Since $(a-1) \geq a/2,\,a \geq 2$, we have
\begin{equation}\label{nlow1a}
\begin{gathered}
P_{\rm e} \geq (2M)^{-1}(2\pi e^{s+\delta})^{-n/2}
\int\limits_{\mbox{\small\boldmath $y$}:\left|
\mbox{\small\boldmath $X$}_{s}(\mbox{\small\boldmath $y$})\right|
\geq 2}\left|\mbox{\boldmath $X$}_{s}(\mbox{\boldmath $y$})\right|\,
d\mbox{\boldmath $y$}\,,
\end{gathered}
\end{equation}
where $\mbox{\boldmath $X$}_{s}(\mbox{\boldmath $y$})$ is defined in
(\ref{defX}). To develop further the right-hand side of
(\ref{nlow1a}) we fix some $r > 0$ and for each
$\mbox{\boldmath $x$}_{i}$ introduce the set
\begin{equation}\label{defZsr}
\begin{gathered}
\mbox{\boldmath $Z$}_{s,r}(i) = \left\{\mbox{\boldmath $y$}:
\left|\|\mbox{\boldmath $y$}\|^{2}-rn\right| \leq \delta n\,,\;
|d(\mbox{\boldmath $y$},\mbox{\boldmath $x$}_{i}) - sn| \leq
\delta n,\,\left|\mbox{\boldmath $X$}_{s}(\mbox{\boldmath $y$})\right|
\geq 2\right\} = \\
= \left\{\mbox{\boldmath $y$}:
\begin{array}{c}
\left|\|\mbox{\boldmath $y$}\|^{2}-rn\right| \leq \delta n\,,\;
|d(\mbox{\boldmath $y$},
\mbox{\boldmath $x$}_{i}) - sn| \leq \delta n \; \mbox{ and} \\
\mbox{there exists $\mbox{\boldmath $x$}_{j}
\neq \mbox{\boldmath $x$}_{i}$ with } |d(\mbox{\boldmath $x$}_{j},
\mbox{\boldmath $y$}) -sn| \leq \delta n
\end{array}
\right\}.
\end{gathered}
\end{equation}
For a measurable set $A \subseteq {\mathbb{R}^{n}}$ denote by $m(A)$
its Lebesque measure. Then
$$
\int\limits_{\mbox{\small\boldmath $y$}:\left|
\mbox{\small\boldmath $X$}_{s}(\mbox{\small\boldmath $y$})\right|
\geq 2}\left|\mbox{\boldmath $X$}_{s}(\mbox{\boldmath $y$})\right|\,
d\mbox{\boldmath $y$} \geq \sum_{i=1}^{M}
m\left(\mbox{\boldmath $Z$}_{s,r}(i)\right)
$$
and from (\ref{nlow1a}) we get

\medskip

{P r o p o s i t i o n \,1}. {\it With any $\delta > 0$ for
the decoding error probability $P_{\rm e}$ the lower bound holds
\begin{equation}\label{glow1}
P_{\rm e}\geq \frac{1}{2M}\,\max_{s,r}
\left\{(2\pi e^{s+\delta})^{-n/2}\sum_{i=1}^{M}
m\left(\mbox{\boldmath $Z$}_{s,r}(i)\right)\right\}\,,
\end{equation}
where $\mbox{\boldmath $Z$}_{s,r}(i)$ is defined in} (\ref{defZsr}).

\medskip

{\bf Example: sphere-packing upper bound.} We show first how to get
the sphere-packing upper bound $E(R,A) \leq E_{\rm sp}(R,A)$ from
(\ref{glow1}) (cf. [1;3, Chapter 7.4]). To simplify
formulas we write below $a \approx b$ if $|a-b| \leq \delta$, where
$\delta = o(1),\,n \to \infty$. Note that
$$
\begin{gathered}
\mbox{\boldmath $Z$}_{s,r}(i) = \mbox{\boldmath $Z$}_{s,r}^{(1)}(i) \setminus
\mbox{\boldmath $Z$}_{s,r}^{(2)}(i)\,, \qquad
\mbox{\boldmath $Z$}_{s,r}^{(1)}(i) = \left\{\mbox{\boldmath $y$}:
\|\mbox{\boldmath $y$}\|^{2}/n \approx r\,,\;
d(\mbox{\boldmath $y$},\mbox{\boldmath $x$}_{i})/n \approx s
\right\}\,,  \\
\mbox{\boldmath $Z$}_{s,r}^{(2)}(i) = \left\{\mbox{\boldmath $y$}:
\|\mbox{\boldmath $y$}\|^{2}/n \approx r\,,\;
d(\mbox{\boldmath $y$},\mbox{\boldmath $x$}_{i})/n \approx s,\,
\left|\mbox{\boldmath $X$}_{s}(\mbox{\boldmath $y$})\right| =
1\right\} = \\
= \left\{\mbox{\boldmath $y$}:
\begin{array}{c}
\|\mbox{\boldmath $y$}\|^{2}/n \approx r\,,\;
d(\mbox{\boldmath $y$},
\mbox{\boldmath $x$}_{i})/n \approx s \; \mbox{ and
there is no $\mbox{\boldmath $x$}_{j}
\neq \mbox{\boldmath $x$}_{i}$ with } d(\mbox{\boldmath $x$}_{j},
\mbox{\boldmath $y$})/n \approx s
\end{array}
\right\}.
\end{gathered}
$$
Then we have
$$
\begin{gathered}
\bigcup_{i=1}^{M}\mbox{\boldmath $Z$}_{s,r}^{(2)}(i) =
\mbox{\boldmath $Y$}_{s} = \left\{\mbox{\boldmath $y$}:
\|\mbox{\boldmath $y$}\|^{2}/n \approx r\,,\;
\left|\mbox{\boldmath $X$}_{s}(\mbox{\boldmath $y$})\right| = 1\right\} = \\
= \left\{\mbox{\boldmath $y$}:
\begin{array}{c}
\|\mbox{\boldmath $y$}\|^{2}/n \approx r
\mbox{ and there exists exactly one
$\mbox{\boldmath $x$}_{i}$ with }
d(\mbox{\boldmath $y$},\mbox{\boldmath $x$}_{i})/n \approx s
\end{array}
\right\}, \\
\mbox{\boldmath $Y$}_{s} \subseteq \mbox{\boldmath $Y$}(r) =
\left\{\mbox{\boldmath $y$}:
\|\mbox{\boldmath $y$}\|^{2}/n \approx r \right\},
\end{gathered}
$$
and the lower bound (\ref{glow1}) takes the form
$$
\begin{gathered}
P_{\rm e}\geq (2M)^{-1}(2\pi e^{s+\delta})^{-n/2}
\left[Mm\left(|\mbox{\boldmath $Z$}_{s,r}^{(1)}(1)|\right) -
m\left(\mbox{\boldmath $Y$}(r)\right)\right]_{+}.
\end{gathered}
$$
The surface area of a $n$-dimensional sphere of radius $a$ is
$S_{n}(a) = n\pi^{n/2}a^{n-1}/\Gamma(n/2+1) \sim
\left(2\pi ea^{2}/n\right)^{n/2}$. Then from a standard geometry we
get
$$
\begin{gathered}
m\left(|\mbox{\boldmath $Z$}_{s,r}^{(1)}(1)|\right) \sim
(2\pi er_{1})^{n/2}\,, \qquad m\left(\mbox{\boldmath $Y$}(r)\right)
\sim (2\pi er)^{n/2}\,, \\
r_{1} = s - \frac{(r-A-s)^{2}}{4A} = r - \frac{(r+A-s)^{2}}{4A}\,.
\end{gathered}
$$
Therefore the lower bound (\ref{glow1}) takes the form
\begin{equation}\label{nlow2a}
\begin{gathered}
P_{\rm e} \gtrsim M^{-1}(e^{s+\delta-1})^{-n/2}
\left[Mr_{1}^{n/2} - r^{n/2}\right]_{+}.
\end{gathered}
\end{equation}

We want to maximize the right-hand side of (\ref{nlow2a}) over
$s,r$. Since we are interested only in exponents in $n$, we may
assume that $Mr_{1}^{n/2} = r^{n/2}$, i.e. $e^{2R}r_{1} = r$.
Then we should maximize the function $f(s,r) = \ln r -s$ provided
$$
s - \frac{(r-A-s)^{2}}{4A} - re^{-2R} = 0\,.
$$
As usual, considering the function
$$
g(s,r) = \ln r -s + \lambda \left[s - \frac{(r-A-s)^{2}}{4A} -
re^{-2R}\right],
$$
and solving the equations $g'_{s} = g'_{r} = 0$, we get
$$
\begin{gathered}
r = \frac{1}{1 -\lambda\left(1-e^{-2R}\right)}\,, \qquad
s = r+A- \frac{2A}{\lambda}\,,
\end{gathered}
$$
where $\lambda$ satisfies the equation
$$
\left(1-e^{-2R}\right)\lambda^{2} + A\left(1-e^{-2R}\right)\lambda -
A = 0 \,.
$$
Therefore
$$
\lambda = \frac{\sqrt{A}}{g_{1}\sqrt{1-e^{-2R}}}\,,
$$
where $g_{1} = g_{1}(R,A)$ is defined in (\ref{sphere1}). Note that
$$
\begin{gathered}
g^{2} - 1 = g\,\sqrt{A\left(1-e^{-2R}\right)}\,, \qquad
1 -\lambda\left(1-e^{-2R}\right) = \frac{1}{g^{2}}\,, \\
\ln r - s = 2\ln g - 1 - A + g\,\sqrt{A\left(1-e^{-2R}\right)}\,.
\end{gathered}
$$
Taking into account that $e^{2R}r_{1} = r$, we get from
(\ref{nlow2a}) and (\ref{sphere1})
$$
\begin{gathered}
\frac{1}{n}\ln \frac{1}{P_{\rm e}} \leq \frac{s-1}{2}- \ln r_{1} =
\frac{s-1}{2} +R - \frac{1}{2}\ln r = \\
= \frac{A-\sqrt{A\left(1-e^{-2R}\right)}g(R,A)}{2} -
\ln g(R,A) +R = E_{\rm sp}(R,A)\,,
\end{gathered}
$$
which gives the sphere-packing upper bound
$E(R,A) \leq E_{\rm sp}(R,A)$.

\medskip

\begin{center}
{\large\bf \S\;3. Lower bound (\ref{glow1}) and code spectrum}
\end{center}

For a code ${\cal C} \subset S^{n-1}(\sqrt{An})$ introduce the code
spectrum function
\begin{equation} \label{codespec}
\begin{gathered}
B(s,t) = \frac{1}{|{\cal C}|} \left|\left\{\mbox{\boldmath $u$},
\mbox{\boldmath $v$} \in {\cal C}: s \leq
\frac{(\mbox{\boldmath $u$},\mbox{\boldmath $v$})}{An} < t \right\}
\right|\,,
\end{gathered}
\end{equation}
and denote
$$
b(\rho,\varepsilon) = \frac{1}{n}\,\ln
B(\rho-\varepsilon,\rho+\varepsilon)\,, \qquad
0 < \varepsilon < \rho\,.
$$

To simplify notation we write below $a \approx b$ if
$|a-b| \leq \delta$, where $\delta = 1/\sqrt{An}$. For some
$r > 0$ we consider only the set of outputs
\begin{equation}\label{defY}
\mbox{\boldmath $Y$}(r) = \left\{\mbox{\boldmath $y$}:
\|\mbox{\boldmath $y$}\|^{2}/n \approx r\right\} \subseteq
{\mathbb{R}}^{n}\,.
\end{equation}

To investigate the function $E(R,A),\,R < R_{\rm crit}$, we use a
variant of the lower bound (\ref{glow1})
\begin{equation}\label{lowsim1}
P_{\rm e} \geq (2M)^{-1}\max_{s,r > 0}\max_{\rho}\left\{
(2\pi e^{s+\delta})^{-n/2}\sum_{i=1}^{M}
m\left(\mbox{\boldmath $Z$}_{s,r}(\rho,i)\right)\right\},
\end{equation}
where
\begin{equation}\label{defZ}
\begin{gathered}
\mbox{\boldmath $Z$}_{s,r}(\rho,i) = \left\{\mbox{\boldmath $y$} \in
\mbox{\boldmath $Y$}(r):
\begin{array}{c}
\mbox{there exists }
\mbox{\boldmath $x$}_{j} \;
%\neq \mbox{\boldmath $x$}_{i} \;
\mbox{with } \rho_{ij} \approx \rho \; \mbox{and } \\
d(\mbox{\boldmath $x$}_{i},\mbox{\boldmath $y$})/n \approx
d(\mbox{\boldmath $x$}_{j},
\mbox{\boldmath $y$})/n \approx s
\end{array}
\right\},
\end{gathered}
\end{equation}
and $\rho_{ij}$ is defined in (\ref{defrho}). We develop the lower
bound (\ref{lowsim1}), relating it to the code spectrum
(\ref{codespec}), i.e. to the distribution of the pairwise inner
products $\{\rho_{ij}\}$.

For codewords $\mbox{\boldmath $x$}_{i},\mbox{\boldmath $x$}_{j}$
with $\rho_{ij} \approx \rho$ introduce the set
\begin{equation}\label{defZ1}
\begin{gathered}
\mbox{\boldmath $Z$}_{s,r}(\rho,i,j) = \left\{\mbox{\boldmath $y$} \in
\mbox{\boldmath $Y$}(r):
\begin{array}{c}
d(\mbox{\boldmath $x$}_{i},\mbox{\boldmath $y$})/n \approx
d(\mbox{\boldmath $x$}_{j},\mbox{\boldmath $y$})/n \approx s
\end{array}
\right\}.
\end{gathered}
\end{equation}
Then for any $i$ from (\ref{defZ}) and (\ref{defZ1}) we have
\begin{equation}\label{defZ2}
\mbox{\boldmath $Z$}_{s,r}(\rho,i) = \bigcup_{j: \rho_{ij} \approx \rho}
\mbox{\boldmath $Z$}_{s,r}(\rho,i,j)\,.
\end{equation}
Denoting
\begin{equation}\label{defzZ}
Z(s,r,\rho) = m\left(\mbox{\boldmath $Z$}_{s,r}(\rho,i,j)\right)
\end{equation}
(since the measure of that set does not depend on indices
$(i,j)$), we have (see Appendix)
\begin{equation}\label{valZ}
\begin{gathered}
\frac{1}{n}\ln Z(s,r,\rho) = \frac{1}{2}\ln \left[2\pi ez(s,r,\rho)
\right] + o(1)\,, \qquad n \to \infty \,,
\end{gathered}
\end{equation}
where
\begin{equation}\label{valZa}
z(s,r,\rho) = r - \frac{(A+r-s)^{2}}{2A(1+\rho)}\,.
\end{equation}

Note that due to (\ref{defZ2}), for the sum in the right-hand side
of (\ref{lowsim1}) for any $\rho$ we have
\begin{equation}\label{asump11}
\begin{gathered}
\sum_{i=1}^{M}m\left(\mbox{\boldmath $Z$}_{s,r}(\rho,i)\right) \leq
\sum_{(i,j): \rho_{ij} \approx \rho}
m\left(\mbox{\boldmath $Z$}_{s,r}(\rho,i,j)\right) =
Z(s,r,\rho)\left|\{(i,j): \rho_{ij} \approx \rho\}\right| = \\
= \exp\left\{\frac{n}{2}\ln \left[2\pi ez(s,r,\rho)
\right] + [R + b(\rho)]n + o(n)\right\}\,,
\end{gathered}
\end{equation}
since for $b(\rho) = b(\rho,\delta)$ the following formula holds
(see (\ref{codespec}))
$$
\left|\{(i,j): \rho_{ij} \approx \rho\}\right| =
e^{Rn}B(\rho-\delta,\rho+\delta) = e^{(R+b(\rho))n}\,.
$$
Suppose that for some $\rho = \rho_{0}$ in the relation
(\ref{asump11}) the following asymptotic equality holds:
\begin{equation}\label{asump1}
\begin{gathered}
\frac{1}{n}\ln\left[\sum_{i=1}^{M}
m\left(\mbox{\boldmath $Z$}_{s,r}(\rho_{0},i)\right)\right] =
\frac{1}{2}\ln \left[2\pi ez(s,r,\rho_{0})
\right] + R + b(\rho_{0}) + o(1)\,, \qquad n \to \infty \,.
\end{gathered}
\end{equation}

Using the functions $s=s(\rho),\,r=r(\rho)$ (they are chosen below),
from (\ref{lowsim1}), (\ref{asump1}) and (\ref{valZa}) for such
$\rho_{0}$ we get
\begin{equation}\label{lowP4}
\begin{gathered}
\frac{1}{n}\ln \frac{1}{P_{\rm e}} \leq \frac{s-1}{2} -\frac{1}{2}
\ln \left[r - \frac{(A+r-s)^{2}}{2A(1+\rho_{0})}\right] -
b(\rho_{0})  + o(1)\,.
\end{gathered}
\end{equation}
We set below
\begin{equation}\label{optsr}
\begin{gathered}
s(\rho) = \frac{A(1-\rho)}{2} +1\,, \qquad
r(\rho) = \frac{A(1+\rho)}{2} +1\,.
\end{gathered}
\end{equation}
Such choice of $s(\rho),r(\rho)$ minimizes (over $s,r$) the
right-hand side of (\ref{lowP4}). Optimality of such $s,r$ can also
be deduced from the formulas (\ref{opt1}) (see Appendix).

For such $s(\rho),r(\rho)$ we have $r- (A+r-s)^{2}/[2A(1+\rho)]= 1$,
and then (\ref{lowP4}) takes the simple form
\begin{equation}\label{lowP4a}
\begin{gathered}
\frac{1}{n}\ln \frac{1}{P_{\rm e}} \leq \frac{A(1-\rho_{0})}{4} -
b(\rho_{0})  + o(1)\,.
\end{gathered}
\end{equation}
Note that $b(\rho) \geq 0$ if there exists a pair
$(\mbox{\boldmath $x$}_{i},\mbox{\boldmath $x$}_{j})$ with
$\rho_{ij} \approx \rho$, and $b(\rho) = - \infty$ if there is no
any pair with $\rho_{ij} \approx \rho$.

We formulate the result obtained as follows.

\medskip

{P r o p o s i t i o n \,2}. {\it If for some $\rho_{0}$ the
condition} (\ref{asump1}) {\it is fulfilled, then the inequality}
(\ref{lowP4a}) {\it for the decoding error probability $P_{\rm e}$
holds}.

\medskip

We show that as such $\rho_{0}$ we may choose the value
$\rho_{0}$, minimizing the right-hand side of (\ref{lowP4a}).
In other words, define $\rho_{0}$ as follows
\begin{equation}\label{defrho0}
\begin{gathered}
A\rho_{0} + 4b(\rho_{0}) = \max_{|\rho| \leq 1}\left\{
A\rho + 4b(\rho)\right\}.
\end{gathered}
\end{equation}

{\it Remark} 3. If there are several such $\rho_{0}$, we may use
any of them. It is not important that we do not know the function
$b(\rho)$. We may use as $b(\rho)$ any lower bound for it
(see proofs of theorems 1 and 2).

\medskip

{P r o p o s i t i o n \,3}. {\it For $\rho_{0}$ from}
(\ref{defrho0}) {\it the condition} (\ref{asump1}) {\it holds and
therefore the inequality} (\ref{lowP4a}) {\it is valid}.

\medskip

{P r o o f}. It is convenient to ``quantize'' the range of possible
values of the normalized inner products $\rho_{ij}$. For that
purpose we partition the whole range $[-1;1]$ of values $\rho_{ij}$
on subintervals of the length $\delta = 1/\sqrt{An}$. There will be
$n_{1} = 2/\delta$ of such subintervals. We may assume that
$\rho_{ij}$ takes values from the set
$\{-1 = \rho_{1} < \ldots < \rho_{n_{1}} = 1\}$.

We call $(\mbox{\boldmath $x$}_{i},\mbox{\boldmath $x$}_{j})$ a
$\rho$-pair if $(\mbox{\boldmath $x$}_{i},
\mbox{\boldmath $x$}_{j})/(An) \approx \rho$. Then $Me^{nb(\rho)}$
is the total number of $\rho$-pairs. We use
$s=s(\rho_{0}),r=r(\rho_{0})$ from (\ref{optsr}) and consider only
outputs $\mbox{\boldmath $y$} \in \mbox{\boldmath $Y$}(r) =
\mbox{\boldmath $Y$}(r(\rho_{0}))$. We say that such a point
$\mbox{\boldmath $y$}$ is $\rho$-covered if there exists a
$\rho$-pair $(\mbox{\boldmath $x$}_{i},\mbox{\boldmath $x$}_{j})$
such that $d(\mbox{\boldmath $x$}_{i},\mbox{\boldmath $y$})/n
\approx d(\mbox{\boldmath $x$}_{j},\mbox{\boldmath $y$})/n \approx
s$. Then the total (taking into account the covering
multiplicities) Lebesque measure of all $\rho$-covered points
$\mbox{\boldmath $y$}$ equals $Me^{nb(\rho)}Z(s,r,\rho)$. \\
Introduce the set $\mbox{\boldmath $Y$}(\rho_{0},\rho)$ of all
$\rho$-covered points $\mbox{\boldmath $y$}$
$$
\mbox{\boldmath $Y$}(\rho_{0},\rho) = \left\{\mbox{\boldmath $y$} \in
\mbox{\boldmath $Y$}(r): \mbox{\boldmath $y$} \mbox{ is }
\rho\!\!-\!\!\mbox{covered}\right\}.
$$
We consider the set $\mbox{\boldmath $Y$}(\rho_{0},\rho)$ and perform its
``cleaning'', excluding from it all points $\mbox{\boldmath $y$}$
that are also $\rho$-covered for any $\rho$ such that
$|\rho - \rho_{0}| \geq 4\delta$, i.e. we consider the set
\begin{equation}\label{defY3a}
\begin{gathered}
\mbox{\boldmath $Y$}'(\rho_{0},\rho_{0}) = \mbox{\boldmath $Y$}(\rho_{0},\rho_{0})
\setminus \bigcup_{|\rho - \rho_{0}| \geq 4\delta}
\mbox{\boldmath $Y$}(\rho_{0},\rho) =  \\
= \left\{\mbox{\boldmath $y$} \in \mbox{\boldmath $Y$}(r):
\begin{array}{c}
\mbox{\boldmath $y$} \mbox{ is } \rho_{0}\!\!-\!\!\mbox{covered and
is not } \rho\!\!-\!\!\mbox{covered} \\
\mbox{for any } \rho \mbox{ such that }
|\rho - \rho_{0}| \geq 4\delta
\end{array}
\right\}.
\end{gathered}
\end{equation}

Each point
$\mbox{\boldmath $y$} \in \mbox{\boldmath $Y$}'(\rho_{0},\rho_{0})$ can
be $\rho$-covered only if $|\rho - \rho_{0}| < 4\delta$. We show
that both sets $\mbox{\boldmath $Y$}(\rho_{0},\rho_{0})$ and
$\mbox{\boldmath $Y$}'(\rho_{0},\rho_{0})$ have essentially the same Lebesque
measures. Note that a $\rho$-pair
$(\mbox{\boldmath $x$}_{i},\mbox{\boldmath $x$}_{j})$ $\rho$-covers
the set $\mbox{\boldmath $Z$}_{s,r}(\rho,i,j)$ from (\ref{defZ1}) with the
Lebesque measure $Z(s,r,\rho)$. We compare the values
$\sum\limits_{|\rho - \rho_{0}| \geq 4\delta}
e^{nb(\rho)}Z(s,r,\rho)$ and $e^{nb(\rho_{0})}Z(s,r,\rho_{0})$
(see (\ref{defY3a})). For that purpose we consider the function
\begin{equation}\label{need8a}
g(\rho) = \frac{1}{n}\ln \frac{e^{nb(\rho)}Z(s,r,\rho)}
{e^{nb(\rho_{0})}Z(s,r,\rho_{0})} = b(\rho) - b(\rho_{0}) +
\frac{1}{2}\ln \frac{z(s,r,\rho)}{z(s,r,\rho_{0})} + o(1)\,,
\end{equation}
where $z(s,r,\rho)$ is defined in (\ref{valZa}). From (\ref{valZa})
we also have
$$
z(s,r,\rho) = 1 + \frac{A(1+\rho_{0})(\rho -\rho_{0})}{2(1+\rho)}\,.
$$
Since $b(\rho) \leq b(\rho_{0}) - A(\rho- \rho_{0})/4$ (see
(\ref{defrho0})), for the function $g(\rho)$ from (\ref{need8a}) we
get
\begin{equation}\label{need8b}
g(\rho) \leq \frac{1}{2}\ln\left[1 + \frac{A(1+\rho_{0})
(\rho - \rho_{0})}{2(1+\rho)}\right] -\frac{A(\rho- \rho_{0})}{4}
\leq -\frac{A(\rho - \rho_{0})^{2}}{4(1+\rho)} \,.
\end{equation}
Since $\rho- \rho_{0}= i\delta\,,\, |i| \geq 4$, after simple
calculations we have
$$
\begin{gathered}
\frac{\sum\limits_{|\rho - \rho_{0}| \geq 4\delta}
e^{nb(\rho)}Z(s,r,\rho)}{e^{nb(\rho_{0})}Z(s,r,\rho_{0})} =
\sum_{|\rho - \rho_{0}| \geq 4\delta}
e^{ng(\rho)} \leq 2\sum_{i \geq 4}
\exp\left\{-\frac{An\delta^{2}i^{2}}{8}\right\} =
2\sum_{i \geq 4}e^{-i^{2}/8} < \frac{1}{2} \,.
\end{gathered}
$$
Therefore we get
$$
e^{nb(\rho_{0})}Z(s,r,\rho_{0}) -
\sum\limits_{|\rho - \rho_{0}| \geq 4\delta}
e^{nb(\rho)}Z(s,r,\rho) > \frac{1}{2}\,
e^{nb(\rho_{0})}Z(s,r,\rho_{0})\,.
$$
Then the total (taking into account the covering multiplicities)
Lebesque measure of all $\rho$-covered points
$\mbox{\boldmath $y$} \in \mbox{\boldmath $Y$}'(\rho_{0},\rho_{0})$ exceeds
$Me^{nb(\rho_{0})}Z(s,r,\rho_{0})/2$. Remind that any point
$\mbox{\boldmath $y$} \in \mbox{\boldmath $Y$}'(\rho_{0},\rho_{0})$ can
be $\rho$-covered only if $|\rho - \rho_{0}| < 4\delta$.

For each point
$\mbox{\boldmath $y$} \in \mbox{\boldmath $Y$}'(\rho_{0},\rho_{0})$ consider
the set $\mbox{\boldmath $X$}_{s}(\mbox{\boldmath $y$})$ defined in
(\ref{defX}), i.e. the set of all codewords
$\{\mbox{\boldmath $x$}_{i}\}$ such that
$d(\mbox{\boldmath $x$}_{i},\mbox{\boldmath $y$})/n \approx s$. The
codewords from $\mbox{\boldmath $X$}_{s}(\mbox{\boldmath $y$})$ satisfy also
the condition $\left|\left(\mbox{\boldmath $x$}_{i},
\mbox{\boldmath $x$}_{j}\right)/(An) - \rho_{0}\right| < 4\delta$,
i.e. the set $\{\mbox{\boldmath $x$}_{i}\}$ constitutes
almost a simplex. It is rather clear that the number
$\left|\mbox{\boldmath $X$}_{s}(\mbox{\boldmath $y$})\right|$ of such
codewords is not exponential on $n$, i.e.
\begin{equation}\label{condX2}
\max_{\mbox{\small\boldmath $y$} \in
\mbox{\boldmath $Y$}'(\rho_{0},\rho_{0})}\left\{\frac{1}{n}
\ln \left|\mbox{\boldmath $X$}_{s}(\mbox{\boldmath $y$})\right|\right\} =
o(1)\,, \qquad n \to \infty\,.
\end{equation}
Formally the validity of (\ref{condX2}) follows from lemma 2
(see below).

Note that if $A_{1},\ldots,A_{N} \subset {\mathbb{R}}^{n}$ are
a measurable sets, and any point $a \in \bigcup\limits_{i}A_{i}$ is
covered by the sets $\{A_{i}\}$ not more than $K$ times, then
\begin{equation}\label{condX3}
m\left(\bigcup_{i=1}^{N}A_{i}\right) \geq \frac{1}{K}\sum_{i=1}^{N}
m(A_{i})\,.
\end{equation}
For $\mbox{\boldmath $y$} \in \mbox{\boldmath $Y$}'(\rho_{0},\rho_{0})$
denote
\begin{equation}\label{defXmax}
\begin{gathered}
\mbox{\boldmath $X$}_{i}(\mbox{\boldmath $y$}) =
\left\{\mbox{\boldmath $x$}_{j}:
\begin{array}{c}
d(\mbox{\boldmath $x$}_{i},\mbox{\boldmath $y$})/n \approx
d(\mbox{\boldmath $x$}_{j}, \mbox{\boldmath $y$})/n \approx s, \;
\rho_{ij} \approx \rho_{0}
\end{array}
\right\}, \\
X_{\rm max} = \max_{i,\mbox{\small\boldmath $y$} \in
\mbox{\boldmath $Y$}'(\rho_{0},\rho_{0})}
\left|\mbox{\boldmath $X$}_{i}(\mbox{\boldmath $y$})\right|.
\end{gathered}
\end{equation}
Due to (\ref{condX2}) we have
\begin{equation}\label{Xmax}
\frac{1}{n}\ln X_{\rm max} = o(1)\,, \quad n \to \infty\,.
\end{equation}
Since any point
$\mbox{\boldmath $y$} \in \mbox{\boldmath $Y$}'(\rho_{0},\rho_{0})$ can be
$\rho$-covered not more than $X_{\rm max}$ times and
$\mbox{\boldmath $Y$}'(\rho_{0},\rho_{0}) \subseteq
\mbox{\boldmath $Y$}(\rho_{0},\rho_{0})$, then from
(\ref{condX2})--(\ref{Xmax}) we get
\begin{equation}\label{asump2}
\begin{gathered}
\frac{1}{n}\ln\left[\sum_{i=1}^{M}
m\left(\mbox{\boldmath $Z$}_{s,r}(\rho_{0},i)\right)\right] \geq
\frac{1}{n}\ln m\left(\mbox{\boldmath $Y$}'(\rho_{0},\rho_{0})\right) \geq \\
\geq \frac{1}{n}\ln\left(Me^{nb(\rho_{0})}Z(s,r,\rho_{0})\right) +
o(1) = \\
= \frac{1}{2}\ln \left[2\pi ez(s,r,\rho_{0})
\right] + R + b(\rho_{0}) + o(1)\,, \qquad n \to \infty \,.
\end{gathered}
\end{equation}
Therefore due to the inequalities (\ref{asump11}) and
(\ref{asump2}), the condition (\ref{asump1}) is fulfilled, and then
the relation (\ref{lowP4a}) holds.

To complete the proof of proposition 2 it remains to establish
the formula (\ref{condX2}). We prove it first for a simpler (but a
more natural) case $\rho^{*} \leq \overline{\tau}_{1}$, and then
consider the general case.

\medskip

{C a s e } $\rho_{0} \leq \overline{\tau}_{1}$. In that case the
relation (\ref{condX2}) follows from simple lemma (see proof in
Appendix).

\medskip

{L e m m a \,1}. {\it Let
$\mbox{\boldmath $y$} \in {\mathbb{R}}^{n}$
with $\|\mbox{\boldmath $y$}\|^{2} = rn$. Let ${\cal C} =
\{\mbox{\boldmath $x$}_{1},\ldots,\mbox{\boldmath $x$}_{M}\}
\subset S^{n-1}(\sqrt{An})$ be a code with
$\|\mbox{\boldmath $x$}_{i} -\mbox{\boldmath $y$} \|^{2} = sn,\,
i=1,\ldots,M$, and $\max\limits_{i \neq j}(\mbox{\boldmath $x$}_{i},
\mbox{\boldmath $x$}_{j}) \leq An\rho$. If
\begin{equation}\label{condX1}
A+r-s \geq 2\sqrt{Ar\rho}\,,
\end{equation}
then} $M \leq 2n$.

\medskip

For $s(\rho),r(\rho)$ from (\ref{optsr}) the condition
(\ref{condX1}) holds, if
\begin{equation}\label{condr}
\rho \leq \frac{A}{2+\sqrt{4+A^{2}}} = \overline{\tau}_{1}(A)\,.
\end{equation}
From lemma 1 and (\ref{condr}) the relation (\ref{condX2}) follows.

\medskip

{G e n e r a l \, c a s e}. Although a code with
$\rho_{0} > \overline{\tau}_{1}$ can
hardly decrease the decoding error probability $P_{\rm e}$, its
investigation needs a bit more efforts. The relation (\ref{condX2})
follows from lemma (see proof in Appendix).

\medskip

{L e m m a \,2}. {\it Let for a code
${\cal C} = \{\mbox{\boldmath $x$}_{1},\ldots,
\mbox{\boldmath $x$}_{M}\} \subset S^{n-1}(\sqrt{An})$ and some
$\rho < 1$ it holds that
$$
\max\limits_{i \neq j}\left|(\mbox{\boldmath $x$}_{i},
\mbox{\boldmath $x$}_{j}) - A\rho n\right| = o(n),\qquad
n \to \infty\,.
$$
Then}  $\ln M = o(n),\,n \to \infty$.

\medskip

It completes the proof of proposition 3. $\qquad \blacktriangle$

\medskip

Using  proposition 3 and two lower bounds for $b(\rho)$ we shall
prove theorems 1 and 2.

\medskip

\begin{center}
{\large\bf \S\;4. Proof of theorem 1}
\end{center}

First we investigate the function $E(R,A)$ for
$0 < R \leq \overline{R}_{1}(A)$ and prove the upper bound
(\ref{exact1a}). Then for
$\overline{R}_{1}(A) < R < R_{\rm crit}(A)$, using the
``straight-line bound'' \cite{SGB1}, we will prove the formula
(\ref{exact1}). To apply  proposition 3 we use the known bound
for the code spectrum. The next result is a slight refinement of
\cite[Theorem 9]{ABL1} (see also \cite[Theorem 1]{Bur4}).

\medskip

{T h e o r e m\; 3}. {\it Let ${\cal C} \subset S^{n-1}(\sqrt{An})$
be a code with $|{\cal C}|=e^{Rn},\,R >0$. Then for any
$\varepsilon = \varepsilon(n) > 0$ there exists $\rho$ such that
$\rho \geq \tau_{R}$ and
\begin{equation} \label{th2.b}
\begin{gathered}
b(\rho) = \frac{1}{n}\,\ln B(\rho-\varepsilon, \rho+\varepsilon)
\geq R - J(t_{R},\rho) + \frac{\ln \varepsilon}{n} + o(1)\,, \quad
n \to \infty \,, \\
J(t,\rho) = (1+2t) \ln \left[2t\rho+ q(t,\rho) \right]
- \ln q(t,\rho) - t\ln [4t(1+t)]  \,, \\
q(t,\rho) = \rho+ \sqrt{(1+2t)^2\rho^2 - 4t(1+t)} \,,
\end{gathered}
\end{equation}
where $t_R, \tau_R$ are defined in} (\ref{defrho}) {\it and}
(\ref{deftau1}), {\it and $o(1)$ does not depend on $\varepsilon$}.

\medskip

Note that
\begin{equation} \label{th2.c}
\begin{gathered}
J'_{\rho}(t,\rho) = \frac{4t(1+t)}
{\rho+ \sqrt{(1+2t)^2\rho^2 - 4t(1+t)}} \,, \\
J''_{\rho \rho}(t,\rho) = -\frac{4t(1+t)}
{[\rho+ \sqrt{(1+2t)^2\rho^2 - 4t(1+t)}]^{2}}\left[1+
\frac{(1+2t)^2\rho}{\sqrt{(1+2t)^2\rho^2 - 4t(1+t)}}\right], \\
J'_{t}(t,\rho) = 2\ln \left[2t\rho+ q(t,\rho)\right] -
\ln[4t(1+t)]\,, \\
\left[R(t) - J(t,\rho)\right]'_{t} =
2\ln \frac{2(1+t)}{2t\rho + q} > 0\,, \quad
J(t_{R},\tau_{R}) = \ln (1+2t_{R})  \,, \quad
J(t_{R},1) = R \,.
\end{gathered}
\end{equation}

\medskip

{P r o p o s i t i o n \,4}. {\it For the function $E(R,A)$ the
upper bound}  (\ref{exact1a}) {\it holds}.

\medskip

{P r o o f}. Due to theorem 2 there exists $\rho \geq \tau_{R}$
such that the inequality (\ref{th2.b}) holds. Denote $\rho^{*}$ the
largest of such $\rho$. Since
$b(\rho_{0}) \geq b(\rho^{*}) - A(\rho_{0}- \rho^{*})/4$
(см. (\ref{defrho0})), from (\ref{lowP4a}) and (\ref{th2.b}) we get
\begin{equation}\label{der0}
\begin{gathered}
\frac{1}{n}\ln \frac{1}{P_{\rm e}} \leq \frac{A(1-\rho_{0})}{4} -
b(\rho_{0})  + o(1) \leq \frac{A(1-\rho^{*})}{4} - b(\rho^{*})  +
o(1) \leq \\
\leq \frac{A(1-\rho^{*})}{4} + J(t_{R},\rho^{*}) - R + o(1) \,.
\end{gathered}
\end{equation}
Note that if $\tau_{R} \leq \overline{\tau}_{1}$ (i.e. if
$R \leq \overline{R}_{1}(A)$) then (see Appendix)
\begin{equation}\label{der1}
\left[J(t_{R},\rho)- A\rho/4\right]'_{\rho} \leq 0\,,
\qquad \rho \geq \tau_{R} \,,
\end{equation}
and therefore the function $J(t_{R},\rho)- A\rho/4$ monotone
decreases on $\rho \geq \tau_{R}$. Since $\rho^{*} \geq \tau_{R}$
then for $\tau_{R} \leq \overline{\tau}_{1}$ we can continue
(\ref{der0}) as follows
\begin{equation}\label{lowP3}
\begin{gathered}
\frac{1}{n}\ln \frac{1}{P_{\rm e}} \leq \frac{A(1-\tau_{R})}{4} +
J(t_{R},\tau_{R}) - R + o(1) = \\
= \frac{A(1-\tau_{R})}{4} + \ln (1+2t_R) - R \,, \qquad
0 < R \leq \overline{R}_{1}\,,
\end{gathered}
\end{equation}
which is the desired upper bound (\ref{exact1a}).
$\qquad \blacktriangle$

\medskip

To prove the relation (\ref{exact1}) note that the best upper bound
for $E(R,A)$ is a combination of the upper bound (\ref{exact1a}) and
the sphere-packing bound via the ``straight-line bound''
\cite{SGB1}, which gives
$$
E(R,A) \leq \frac{A(1-\overline{\tau}_{1})}{4} +
\ln (1+2\overline{t}_{1}) - R\,, \quad
\overline{R}_{1} \leq R \leq R_{\rm crit}\,.
$$
On the other hand, the random coding bound \cite{Sh,G1} gives
$$
E(R,A) \geq E_{\rm sp}(R_{\rm crit},A) + R_{\rm crit} - R\,, \qquad
R \leq R_{\rm crit}\,,
$$
where $E_{\rm sp}(R,A)$ is defined in (\ref{sphere1}). Together with
the formula (\ref{check1}) it completes the proof of theorem 1.
$\qquad \blacktriangle$

\begin{center}
{\large\bf \S\;5. Proof of theorem 2}
\end{center}

As was already mentioned in \S\,1, for $R > 0.234$ the upper
bounds for the minimal code distance \cite[p. 20]{KL} of a spherical
code and its spectrum \cite{BHL1} can be improved, if the linear
programming bound is not directly applied to the original spherical
code, but to its subcodes on spherical caps. The same approach
allows to improve the upper bound for $E(R,A)$ as well. For that
purpose we will need a bound for a code spectrum better than
(\ref{th2.b}). The bound obtained
below (theorem 4), probably, is equivalent to the similar bound in
\cite[Theorem 3]{BHL1} (expressed in a different terms), but its
derivation is simpler and a more accurate.

Since we are interested only in angles between codewords
$\mbox{\boldmath $x$}_{i},\mbox{\boldmath $x$}_{j}$, for the
formulas simplification we may set $An = 1$, and consider a code
${\cal C} \subset S^{n-1}(1) = S^{n-1}$. Let
$T^{n}_{\theta}(\mbox{\boldmath $z$})$ be the {\it spherical cap}
with half-angle $0 \leq \theta \leq \pi/2$ and center
$\mbox{\boldmath $z$} \in S^{n-1}$, i.e.
$$
T^{n}_{\theta}(\mbox{\boldmath $z$}) = \left\{\mbox{\boldmath $x$}
\in S^{n-1}: (\mbox{\boldmath $x$},\mbox{\boldmath $z$}) \geq
\cos \theta\right\}.
$$
It will be convenient to consider subcodes of ${\cal C}$ not on
spherical caps $T^{n}_{\theta}(\mbox{\boldmath $z$})$, but on
related with them thin ring-shaped surfaces
$D^{n}_{\theta}(\mbox{\boldmath $z$})$. We set further
$\delta = 1/n^{2}$, and denote
$D^{n}_{\theta}(\mbox{\boldmath $z$})$ as
\begin{equation}\label{defD1}
D^{n}_{\theta}(\mbox{\boldmath $z$}) =
T^{n}_{\theta}(\mbox{\boldmath $z$}) \setminus
T^{n}_{\theta - \delta}(\mbox{\boldmath $z$}) =
\left\{\mbox{\boldmath $x$} \in S^{n-1}: \cos\theta \leq
(\mbox{\boldmath $x$},\mbox{\boldmath $z$}) \leq
\cos(\theta - \delta)\right\}.
\end{equation}
Denote $D_{n}(\theta)$ the surface area of
$D^{n}_{\theta}(\mbox{\boldmath $z$})$. Then \cite[formula (21)]{Sh}
$$
D_{n}(\theta) = \frac{(n-1)\pi^{(n-1)/2}}{\Gamma((n+1)/2)}
\int\limits_{\theta-\delta}^{\theta}\sin^{n-2}u\,du\,,  \qquad
\delta \leq \theta \leq \pi/2\,.
$$
It is not difficult to show that
$$
1 - \frac{1}{2n\sin\theta} \leq
\frac{D_{n}(\theta)\Gamma((n+1)/2)n^{2}}
{\pi^{(n-1)/2}(n-1)\sin^{n-2}\theta} \leq 1\,.
$$
Since the surface area $|S^{n-1}|$ of the sphere $S^{n-1}$ equals
$n\pi^{n/2}/\Gamma(n/2+1)$, we have uniformly over
$1/n \leq \theta \leq \pi/2$
$$
\frac{1}{n}\ln \frac{D_{n}(\theta)}{|S^{n-1}|} =
\ln \sin \theta + o(1)\,, \qquad n \to \infty\,.
$$

For the code ${\cal C} \subset S^{n-1}$ and $\theta$ such that
$\max\{\arcsin e^{-R},1/n\} \leq \theta \leq \pi/2$, and
$\mbox{\boldmath $z$} \in S^{n-1}$ we consider the subcode
${\cal C}(\theta,\mbox{\boldmath $z$}) =
{\cal C} \cap D^{n}_{\theta}(\mbox{\boldmath $z$})$ with
$|{\cal C}(\theta,\mbox{\boldmath $z$})| =
e^{nr(\mbox{\small\boldmath $z$})}$ codewords. Then
$$
\frac{1}{m(S^{n-1})}
\int\limits_{\mbox{\small\boldmath $z$} \in S^{n-1}}
|{\cal C}(\theta,\mbox{\boldmath $z$})|\,d\mbox{\boldmath $z$} =
\frac{|{\cal C}|D_{n}(\theta)}{|S^{n-1}|} =
\exp\left\{(R + \ln\sin\theta)n + o(n)\right\},
$$
i.e. in average (over $\mbox{\boldmath $z$} \in S^{n-1}$)
a subcode ${\cal C}(\theta,\mbox{\boldmath $z$})$ has the
rate $r = R + \ln\sin\theta + o(1)$. All its
$|{\cal C}(\theta,\mbox{\boldmath $z$})|$ codewords are located in
the ball $B^{n}(\sin\theta,\mbox{\boldmath $z$}')$ of radius
$\sin\theta$ and centered at
$\mbox{\boldmath $z$}' = \mbox{\boldmath $z$}\cos\theta$.
Moreover, they are located in a thin (of thickness $\sim \delta$)
torus orthogonal to $\mbox{\boldmath $z$}$. If
$\mbox{\boldmath $x$} \in D^{n}_{\theta}(\mbox{\boldmath $z$})$,
then we denote $\mbox{\boldmath $x$}' = \mbox{\boldmath $x$} -
\mbox{\boldmath $z$}'$ the corresponding vector from
$B^{n}(\sin\theta,\mbox{\boldmath $z$}')$. The original angle
$\varphi$ between two vectors $\mbox{\boldmath $x$},
\mbox{\boldmath $y$} \in D^{n}_{\theta}(\mbox{\boldmath $z$})$
becomes the angle $\varphi' + O(\delta)$ between the vectors
$\mbox{\boldmath $x$}',\mbox{\boldmath $y$}' \in
B^{n}(\sin\theta,\mbox{\boldmath $z$}')$, where
$\sin(\varphi'/2) = \sin(\varphi/2)/\sin\theta$. The original value
$\rho = \cos\varphi$ becomes the value $\rho' + O(\delta)$, where
$\rho' = \cos\varphi'$ is defined by the formula
\begin{equation} \label{rho'}
1-\rho = (1-\rho')\sin^{2}\theta \,,
\end{equation}
since
$$
\begin{gathered}
\rho' = \cos\left(2\arcsin\left(\frac{\sin(\varphi/2)}{\sin\theta}
\right)\right) = 1-\frac{2\sin^{2}(\varphi/2)}{\sin^{2}\theta} =
1-(1-\rho)e^{2(R-r)}\,.
\end{gathered}
$$
The angle $\varphi'$ and the value $\rho'$ correspond to the case
when the vectors $\mbox{\boldmath $x$}',\mbox{\boldmath $y$}'$
are orthogonal to $\mbox{\boldmath $z$}$.
The code ${\cal C}(\theta,\mbox{\boldmath $z$})$ is then transferred
to the code ${\cal C}'(\mbox{\boldmath $z$}) =
{\cal C}'(\theta,\mbox{\boldmath $z$}) \subset
B^{n}(\sin\theta,\mbox{\boldmath $z$}')$.

To evaluate the average number $e^{nb_{{\cal C}}(\rho)}$ of
$\rho$-neighbors in the code ${\cal C}$, we consider any pair
$\mbox{\boldmath $x$}_{i},\mbox{\boldmath $x$}_{j}$ with
$(\mbox{\boldmath $x$}_{i},\mbox{\boldmath $x$}_{j}) = \rho$ and
introduce the sets
$$
\begin{gathered}
\mbox{\boldmath $Z$}(\mbox{\boldmath $x$},a) = \left\{
\mbox{\boldmath $z$} \in S^{n-1}: (\mbox{\boldmath $x$},
\mbox{\boldmath $z$}) \geq a\right\}\,, \\
\mbox{\boldmath $Z$}(\mbox{\boldmath $x$},\mbox{\boldmath $y$},a) =
\left\{\mbox{\boldmath $z$} \in S^{n-1}: (\mbox{\boldmath $x$},
\mbox{\boldmath $z$}) \geq a\,\mbox{ and }\,
(\mbox{\boldmath $y$},\mbox{\boldmath $z$}) \geq a\right\}\,.
\end{gathered}
$$
Denote by $\Omega_{n}(\theta)$ the surface area of the spherical cap
$T^{n}_{\theta}(\mbox{\boldmath $z$})$. For $0 \leq \theta < \pi/2$
we have
$$
\Omega_{n}(\theta) = \frac{\pi^{(n-1)/2}\sin^{n-1}\theta}
{\Gamma((n+1)/2)\cos \theta}\left(1+o(1)\right), \qquad
n \to \infty\,.
$$
Then for the Lebesque measure $m(a)$ of the set
$\mbox{\boldmath $Z$}(\mbox{\boldmath $x$},a)$ we have
$$
m(a) = m\left(\mbox{\boldmath $Z$}(\mbox{\boldmath $x$},a)\right)
= \Omega_{n}(\arccos a)\,.
$$
We evaluate the Lebesque measure $m(\rho,a)$ of the set
$\mbox{\boldmath $Z$}(\mbox{\boldmath $x$},\mbox{\boldmath $y$},a)$
provided $(\mbox{\boldmath $x$},\mbox{\boldmath $y$}) = \rho$.
Note that if $\mbox{\boldmath $x$},\mbox{\boldmath $y$} \in S^{n-1}$
and $(\mbox{\boldmath $x$},\mbox{\boldmath $y$}) = \rho$, then
$\|\mbox{\boldmath $x$} + \mbox{\boldmath $y$}\|^{2} = 2(1+\rho)$.
Therefore $\mbox{\boldmath $v$} = \\
(\mbox{\boldmath $x$} + \mbox{\boldmath $y$})/\sqrt{2(1+\rho)} \in
S^{n-1}$, and then
$$
\begin{gathered}
\mbox{\boldmath $Z$}(\mbox{\boldmath $x$},\mbox{\boldmath $y$},a)
\subseteq \left\{\mbox{\boldmath $z$} \in S^{n-1}:
(\mbox{\boldmath $x$} + \mbox{\boldmath $y$},
\mbox{\boldmath $z$}) \geq 2a\right\} = \\
= \left\{\mbox{\boldmath $z$} \in S^{n-1}: (\mbox{\boldmath $v$},
\mbox{\boldmath $z$}) \geq a\sqrt{2/(1+\rho)}\right\} =
\mbox{\boldmath $Z$}\left(\mbox{\boldmath $v$},
a\sqrt{2/(1+\rho)}\right)\,.
\end{gathered}
$$
Therefore we get
$$
m(\rho,a) = m\left(\mbox{\boldmath $Z$}(\mbox{\boldmath $x$},
\mbox{\boldmath $y$},a)\right) \leq
m\left(\mbox{\boldmath $Z$}\left(\mbox{\boldmath $v$},
a\sqrt{2/(1+\rho)}\right)\right) = \Omega_{n}
\left(\arccos\left(a\sqrt{2/(1+\rho)}\right)\right)\,.
$$
That upper bound for $m(\rho,a)$ is logarithmically
(as $n \to \infty$) exact. In particular, if $a = \cos \theta$ and
$(\mbox{\boldmath $x$},\mbox{\boldmath $y$}) = \rho$, then
$$
\begin{gathered}
\frac{1}{n}\ln \frac{m(\cos \theta)}{m(\rho,\cos \theta)} \geq
\ln\sin\theta - \ln\sin\left(\arccos\left(\sqrt{2/(1+\rho)}\,
\cos\theta\right)\right) = \\
= \ln\sin\theta - \ln\sqrt{1-2\cos^{2}\theta/(1+\rho)}\,.
\end{gathered}
$$

We use below the values $\rho' = \rho'(\rho,\theta)$ from
and (\ref{rho'}) and $\varepsilon' = \varepsilon/\sin^{2}\theta$.
Then denoting $B_{{\cal C}}(\rho) =
B_{{\cal C}}(\rho-\varepsilon,\rho+\varepsilon),\,
B_{{\cal C}'(\mbox{\small\boldmath $z$})}(\rho') =
B_{{\cal C}'(\mbox{\small\boldmath $z$})}
(\rho'-\varepsilon',\rho'+\varepsilon')$, for any
$\rho, \varepsilon$ we have
\begin{equation}\label{edge1}
B_{{\cal C}}(\rho)|{\cal C}| = \frac{1}{m(\rho,\cos\theta)}
\int\limits_{\mbox{\small\boldmath $z$} \in S^{n-1}}
B_{{\cal C}'(\mbox{\small\boldmath $z$})}(\rho')
|{\cal C}'(\mbox{\boldmath $z$})|\,d\mbox{\boldmath $z$}\,.
\end{equation}
Indeed, the value $B_{{\cal C}}(\rho)|{\cal C}|$ is the total number
of pairs
$\mbox{\boldmath $x$}_{i},\mbox{\boldmath $x$}_{j} \in {\cal C}$
with $|(\mbox{\boldmath $x$}_{i},\mbox{\boldmath $x$}_{j}) - \rho|
\leq \varepsilon$, and $B_{{\cal C}'(\mbox{\small\boldmath $z$})}
(\rho')|{\cal C}'(\mbox{\boldmath $z$})|$ is the total number of
similar pairs
$\mbox{\boldmath $x$}'_{i},\mbox{\boldmath $x$}'_{j} \in
{\cal C}'(\mbox{\boldmath $z$})$ with $|(\mbox{\boldmath $x$}_{i}',
\mbox{\boldmath $x$}_{j}')/(\|\mbox{\boldmath $x$}_{i}'\|\cdot
\|\mbox{\boldmath $x$}_{j}'\|)
-\rho'| \leq \varepsilon'$. Moreover, each pair
$\mbox{\boldmath $x$}_{i}',\mbox{\boldmath $x$}_{j}'
\in {\cal C}'(\mbox{\boldmath $z$})$  gives the contribution
$m(\rho,\cos\theta)$ to the integral, from which the formula
(\ref{edge1}) follows. From (\ref{edge1}) for any set
${\cal A} \subseteq S^{n-1}$ we have
\begin{equation}\label{edge1a}
e^{nb_{{\cal C}}(\rho)} \geq \frac{1}{m(\rho,\cos\theta)|{\cal C}|}
\int\limits_{\mbox{\small\boldmath $z$} \in {\cal A}}
e^{nb_{{\cal C}'(\mbox{\small\boldmath $z$})}(\rho')}
|{\cal C}'(\mbox{\boldmath $z$})|\,d\mbox{\boldmath $z$} \,,
\end{equation}
and also
$$
|{\cal C}| = \frac{1}{m(\cos\theta)}
\int\limits_{\mbox{\small\boldmath $z$} \in S^{n-1}}
|{\cal C}'(\mbox{\boldmath $z$})|\,d\mbox{\boldmath $z$} \geq
\frac{1}{m(\cos\theta)}
\int\limits_{\mbox{\small\boldmath $z$} \in {\cal A}}
|{\cal C}'(\mbox{\boldmath $z$})|\,d\mbox{\boldmath $z$}\,.
$$
The code ${\cal C}'(\mbox{\boldmath $z$})$ has the rate
$r(\mbox{\boldmath $z$}) =
(\ln |{\cal C}'(\mbox{\boldmath $z$})|)/n$. Then there exists
$r_{0}$ such that
\begin{equation}\label{edge2}
\begin{gathered}
|{\cal C}| = \frac{e^{o(n)}}{m(\cos\theta)}\max_{t} \left\{e^{tn}
m\left(\mbox{\boldmath $z$} \in S^{n-1}: |r(\mbox{\boldmath $z$}) -
t| \leq \varepsilon\right)\right\} = \frac{e^{r_{0}n+o(n)}m(S_{0})}
{m(\cos\theta)}\,, \\
S_{0} = \left\{\mbox{\boldmath $z$} \in S^{n-1}:
|r(\mbox{\boldmath $z$}) - r_{0}| \leq \varepsilon\right\}\,.
\end{gathered}
\end{equation}
Since $m(S_{0}) \leq m(S^{n-1})$ then
\begin{equation}\label{edge2a}
r_{0} \geq \frac{1}{n}
\ln \frac{|{\cal C}|m(\cos\theta)}{m(S^{n-1})} =
R + \ln \sin\theta + o(1)\,.
\end{equation}
We set ${\cal A} = S_{0}$ and $\varepsilon = o(1),\, n \to \infty$.
Then using the Jensen inequality, from (\ref{edge1a}) and
(\ref{edge2}) we have
$$
\begin{gathered}
e^{nb_{{\cal C}}(\rho)} \geq
\frac{1}{m(\rho,\cos\theta)|{\cal C}|}
\int\limits_{\mbox{\small\boldmath $z$} \in S_{0}}
e^{nb_{{\cal C}'(\mbox{\small\boldmath $z$})}(\rho')}
|{\cal C}'(\mbox{\boldmath $z$})|\,d\mbox{\boldmath $z$} \geq \\
\geq \frac{m(\cos\theta)e^{o(n)}}
{m(\rho,\cos\theta)m(S_{0})}
\int\limits_{\mbox{\small\boldmath $z$} \in S_{0}}
e^{nb_{{\cal C}'(\mbox{\small\boldmath $z$})}(\rho')}\,
d\mbox{\boldmath $z$} \geq \\
\geq \frac{m(\cos\theta)e^{o(n)}}{m(\rho,\cos\theta)}\exp\left\{
\frac{n}{m(S_{0})}\int\limits_{\mbox{\small\boldmath $z$} \in S_{0}}
b_{{\cal C}'(\mbox{\small\boldmath $z$})}(\rho')\,
d\mbox{\boldmath $z$}\right\},
\end{gathered}
$$
from which we get
\begin{equation}\label{edge3a}
\begin{gathered}
b_{{\cal C}}(\rho) \geq \frac{1}{n}\ln\frac{m(\cos\theta)}
{m(\rho,\cos\theta)} + \frac{1}{m(S_{0})}
\int\limits_{\mbox{\small\boldmath $z$} \in S_{0}}
b_{{\cal C}'(\mbox{\small\boldmath $z$})}(\rho')\,
d\mbox{\boldmath $z$} + o(1)\,.
\end{gathered}
\end{equation}
Due to theorem 3 for each code
${\cal C}'(\mbox{\boldmath $z$}),\,\mbox{\boldmath $z$} \in S_{0}$,
there exists $\rho'' = \rho''(\mbox{\boldmath $z$})$ such that
$\rho'' \geq \tau_{r_{0}}$ and
$$
\begin{gathered}
b_{{\cal C}'(\mbox{\small\boldmath $z$})}(\rho'') \geq r_{0} -
J(t_{r_{0}},\rho'') + o(1)\,.
\end{gathered}
$$
Therefore there exists $\rho' \geq \tau_{r_{0}}$ and the
corresponding $\rho = \rho(\rho')$ from (\ref{rho'}) such that from
the inequality (\ref{edge3a}) we get
\begin{equation} \label{th2.f}
\begin{gathered}
b_{{\cal C}}(\rho) \geq \frac{1}{n}\ln\frac{m(\cos\theta)}
{m(\rho,\cos\theta)} + r_{0} - J(t_{r_{0}},\rho') + o(1) \geq  \\
= \frac{1}{n}\ln\frac{m(\cos\theta)}{m(\rho,\cos\theta)} + R +
\ln \sin \theta - J(t_{R + \ln \sin \theta},\rho') + o(1) \geq \\
\geq R+ 2\ln\sin\theta - J(t_{R + \ln\sin\theta},\rho') -
\ln\sqrt{1-2\cos^{2}\theta/(1+\rho)} + o(1) = \\
= R+ \ln\sin\theta - J(t_{R + \ln\sin\theta},\rho') +
\frac{1}{2}\ln\frac{(1+\rho)}{(1+\rho')} + o(1) \,,
\end{gathered}
\end{equation}
where we used the formula (\ref{edge2a}) and monotonicity of the
function $r - J(t_{r},\rho)$ on $r$ (see (\ref{th2.c})),  and
$\rho' = \rho'(\rho,\theta)$ is defined in (\ref{rho'}). After the
variable change $\sin\theta = e^{r-R}$ from (\ref{th2.f}) we get

\medskip

{T h e o r e m\; 4}. {\it Let ${\cal C} \subset S^{n-1}(1)$ be a
code with $|{\cal C}|=e^{Rn},\,R >0$. Then for any $r \leq R$ there
exists $\rho'$ such that $\rho' \geq \tau_{r}$ and for
$\rho = 1-(1-\rho')e^{2(r-R)}$ the following inequality holds}
\begin{equation}\label{th2.e1}
b_{{\cal C}}(\rho) \geq r - J(t_{r},\rho') +
\frac{1}{2}\ln\frac{(1+\rho)}{(1+\rho')} + o(1)\,.
\end{equation}

Using the relation (\ref{th2.e1}) in the inequality (\ref{lowP4a})
we prove theorem 2. We have
\begin{equation}\label{imp0}
\begin{gathered}
\frac{1}{n}\ln \frac{1}{P_{\rm e}} \leq \min_{r \leq R}
\max_{\rho' \geq \tau_{r}}\left\{\frac{A(1-\rho)}{4} - b(\rho)
\right\} + o(1) \leq \\
\leq \min_{r \leq R}\max_{\rho' \geq \tau_{r}}
\left\{\frac{A(1-\rho')e^{2(r-R)}}{4} - r + J(t_{r},\rho') +
\frac{1}{2}\ln\frac{1+\rho'}{1+\rho}\right\} =
\min_{r \leq R}\max_{\rho \geq \tau_{r}} f(r,\rho)\,,
\end{gathered}
\end{equation}
where
$$
f(r,\rho) = \frac{A(1-\rho)e^{2(r-R)}}{4} + R- 2r + J(t_{r},\rho)
+ \frac{1}{2}\ln\frac{1+\rho}{2e^{2(R-r)}+\rho-1}\,.
$$
With $t = t_{r}$ and $(1-\tau_{r})e^{2(r-R)} = 2z$ we have
$$
\begin{gathered}
f'_{\rho} = -\frac{Ae^{2(r-R)}}{4} - \frac{1}{2(2e^{2(R-r)}+\rho-1)}
+ \frac{4t(1+t)}{\rho+ \sqrt{(1+2t)^2\rho^2 - 4t(1+t)}} +
\frac{1}{2(1+\rho)}\,, \\
f'_{\rho}\big|_{\rho = \tau_{r}} = -\frac{Ae^{2(r-R)}}{4} -
\frac{1}{2(2e^{2(R-r)}+\tau_{r}-1)} + \frac{1}{2(1-\tau_{r})} = \\
= \frac{Az^{2} - (A+2)z + 1}
{2(1-z)(1-\tau_{r})}\,,
\qquad f''_{\rho \rho}  < 0\,.
\end{gathered}
$$
Since $f''_{\rho \rho} < 0$ then $\rho = \tau_{r}$ is optimal if
$f'_{\rho}\big|_{\rho = \tau_{r}} \leq 0$. Since $r \leq R$ then
$z \leq 1$. Therefore $f'_{\rho}\big|_{\rho = \tau_{r}} \leq 0$ if
the following inequalities are fulfilled:
\begin{equation}\label{condf}
\frac{2}{A+2+\sqrt{A^{2}+4}} \leq z \leq
\frac{A+2+\sqrt{A^{2}+4}}{2A}\,.
\end{equation}
The right one of the inequalities (\ref{condf}) is always satisfied.
The left one of the inequalities (\ref{condf}) is equivalent to the
inequality
\begin{equation}\label{condf1}
f_{2}(r) = 2r + \ln(1-\tau_{r}) \geq 2R - 2R_{\rm crit}(A)\,.
\end{equation}
The next simple technical lemma concerns the function $f_{2}(r)$ in
the left-hand side of (\ref{condf1}).

\medskip

{L e m m a \,3}. {\it The function $f_{2}(r)$ from} (\ref{condf1})
{\it monotone decreases on $0 \leq r < \overline{R}_{2}$, and
monotone increases on $r > \overline{R}_{2}$, where
$\overline{R}_{2}$ is defined in} (\ref{defR2}). {\it Moreover, the
formula holds}
\begin{equation}\label{condf3}
\ln\left(1- \overline{\tau}_{1}(A)\right) = -2R_{\rm crit}(A)\,,
\qquad A > 0\,.
\end{equation}

\medskip

Since the function $E(R,A),\,R \geq \overline{R}_{1}(A)$, is known
exactly (see theorem 1), we consider only the case
$R < \overline{R}_{1}(A)$. Then two cases are possible:
$R \leq \min\{\overline{R}_{1}(A),\overline{R}_{2}\}$ and
$\overline{R}_{2} < R < \overline{R}_{1}(A)$.

\medskip

{C a s e } $R \leq \min\{\overline{R}_{1}(A),\overline{R}_{2}\}$.
For $R \leq \overline{R}_{2}$ minimum (over $r \leq R$) in the
left-hand side of (\ref{condf1}) is attained when $r = R$, and then
due to (\ref{condf3}) the inequality (\ref{condf1}) reduces to the
condition $\tau_{R} \leq \overline{\tau}_{1}(A)$, i.e. to
$R \leq \overline{R}_{1}(A)$. Therefore if
$r \leq R \leq \min\{\overline{R}_{1}(A),\overline{R}_{2}\}$ then
the inequalities (\ref{condf1}) and (\ref{condf}) are fulfilled, and
then $\rho = \tau_{r}$ is optimal in the right-hand side of
(\ref{imp0}). Since
$J(t_{r},\tau_{r}) = \ln (1+2t_{r}) = -\ln(1-\tau_{r}^{2})/2$
(see (\ref{th2.c}) and (\ref{deftau1})), then (\ref{imp0}) takes the
form
\begin{equation}\label{imp01}
\frac{1}{n}\ln \frac{1}{P_{\rm e}} \leq
\min_{r \leq R}f(r,\tau_{r}) = \min_{r \leq R}
C(v(r)) - R\,, \qquad R \leq
\min\{\overline{R}_{1}(A),\overline{R}_{2}\}\,,
\end{equation}
where
\begin{equation}\label{imp02}
\begin{gathered}
C(v) = \frac{Av}{4} - \frac{1}{2}\ln[v(2 - v)]\,,
\qquad v(r) = (1-\tau_{r})e^{2(r-R)}\,.
\end{gathered}
\end{equation}
Note that for $r = R$ the inequality (\ref{imp01}) reduces to the
previous bound (\ref{exact1a}). We show that such $r$ is optimal
in (\ref{imp01}). We have
$$
\begin{gathered}
4v(2-v)C'_{v} = -Av^{2}+2(A+2)v - 4 \,, \qquad C''_{v^{2}} > 0\,.
\end{gathered}
$$
Since $0 \leq v \leq 1\,$, the equation $C'_{v} = 0$ has the unique
root $v_{1}$, where
\begin{equation}\label{defv1}
v_{1} = \frac{4}{A+2+\sqrt{A^{2}+4}} = e^{-2R_{\rm crit}(A)}\,.
\end{equation}
The function $C(v),\,0 \leq v \leq 1$, monotone decreases on
$0 \leq v < v_{1}$ and monotone increases on $v > v_{1}$. Note that
since $v(r) = e^{f_{2}(r)-2R}$, then (see lemma 3) the function
$v(r)$ monotone decreases on $0 \leq r < \overline{R}_{2}$ and
monotone increases on $r > \overline{R}_{2}$.

If now $R \leq \min\{\overline{R}_{1}(A),\overline{R}_{2}\}$,
then $v(r) \geq v_{1}$ for $r \leq R$. Therefore $r = R$ is optimal
in (\ref{imp01}), and then (\ref{imp01}) reduces to the
previous bound (\ref{exact1a}).

\medskip

{C a s e } $\overline{R}_{2} < R < \overline{R}_{1}(A)$ (i.e.
$A > A_{0}$). Then
$\overline{R}_{2} < \overline{R}_{3}(A) < \overline{R}_{1}(A)$,
where $\overline{R}_{3}(A)$ is defined in (\ref{defR2}). Consider
first the case $\overline{R}_{2} \leq R \leq \overline{R}_{3}(A)$.
It is simple to check that then the inequality (\ref{condf1}) is
again satisfied (see (\ref{defR2})). Therefore $\rho = \tau_{r}$
is optimal in the right-hand side of (\ref{imp0}), and (\ref{imp0})
takes the form (\ref{imp01}). Since $R \leq \overline{R}_{3}(A)$,
then $v(r) \geq v_{1}$ for $r \leq R$. Since
$R \geq \overline{R}_{2}$ then $r = \overline{R}_{2}$ is optimal in
(\ref{imp01}), and then from (\ref{imp01}) the second of bounds
(\ref{exact1b}) follows.

It remains to consider the case $\overline{R}_{2} \leq
\overline{R}_{3}(A) \leq R \leq \overline{R}_{1}(A)$. Since minimum
of $C(v)$ over $0 \leq v \leq 1$ is attained for $v = v_{1}$
(see (\ref{defv1})), then
\begin{equation}\label{minC}
\begin{gathered}
\min_{0 \leq v \leq 1} C(v) = C(v_{1}) =
E_{\rm sp}(R_{\rm crit},A) + R_{\rm crit} \,,
\end{gathered}
\end{equation}
where the formula was used
$$
E_{\rm sp}(R_{\rm crit},A) + R_{\rm crit} = \frac{Av_{1}}{4} -
\frac{1}{2}\ln v_{1} - \frac{1}{2}\ln(2 - v_{1})\,.
$$
Now in the right-hand side of (\ref{imp0}) we set $r$ such that
$v(r) = v_{1}$ (it is possible when $R \geq \overline{R}_{3}$). Then
again the inequality (\ref{condf1}) is fulfilled and
$\rho = \tau_{r}$ is optimal in the right-hand side of (\ref{imp0}).
From (\ref{imp01}) and (\ref{minC}) the first of upper bounds
(\ref{exact11}) follows. The upper bound (\ref{exact11}) can also be
proved applying the ``straight-line bound'' to the sphere-packing
bound and the second of upper bounds (\ref{exact1b}) at
$R = \overline{R}_{3}$, and the formula
$$
E_{\rm sp}(R_{\rm crit},A) + R_{\rm crit} - \overline{R}_{3} =
\frac{Aae^{-2\overline{R}_{3}}}{4}-
\dfrac{1}{2}\ln(2 - ae^{-2\overline{R}_{3}}) - \dfrac{1}{2}\ln a\,,
$$
which is simple to check using the relations (\ref{userel1}).
It completes the proof of theorem 2. $\qquad \blacktriangle$

\medskip

%\newpage

\hfill {\large\sl APPENDIX}

\medskip

{P r o o f \, o f \, f o r m u l a \, (\ref{valZ}).} Without loss of
generality we may assume that $\mbox{\boldmath $x$}_{i},
\mbox{\boldmath $x$}_{j},\mbox{\boldmath $y$}$ have the form
$$
\begin{gathered}
\mbox{\boldmath $x$}_{i} = (x_{1},x_{2},0,\dots,0), \quad
\mbox{\boldmath $x$}_{j} = (-x_{1},x_{2},0,\dots,0), \quad
\mbox{\boldmath $y$} = (0,y_{2},y_{3},\dots,y_{n}),
\end{gathered}
$$
from which we have
$$
\begin{gathered}
d(\mbox{\boldmath $x$}_{i},\mbox{\boldmath $x$}_{j}) =
4x_{1}^{2} = 2An(1-\rho_{ij})\,, \\
d(\mbox{\boldmath $x$}_{i},\mbox{\boldmath $y$}) = x_{1}^{2} +
(y_{2}-x_{2})^{2} + \sum_{k=3}^{n}y_{k}^{2} = sn \,, \\
x_{1}^{2} + x_{2}^{2} = An\,, \qquad
\sum_{k=2}^{n}y_{k}^{2} = rn \,.
\end{gathered}
$$
Solving those equations we get
\begin{equation}\label{opt1}
x_{1} = \sqrt{\frac{An(1-\rho_{ij})}{2}}\,, \qquad
x_{2} =\sqrt{\frac{An(1+\rho_{ij})}{2}}\,, \qquad
y_{2} = \frac{(A+r-s)n}{\sqrt{2An(1+\rho_{ij})}}\,,
\end{equation}
and therefore
$$
\sum_{k=3}^{n}y_{k}^{2} = rn - y_{2}^{2} = rn -
\frac{(A+r-s)^{2}n}{2A(1+\rho_{ij})} = r_{1}n\,,
$$
from which the formula (\ref{valZ}) follows.
$\qquad \blacktriangle$

Optimality of $s(\rho),r(\rho)$ from the formulas (\ref{optsr})
also follows from (\ref{opt1}).

\medskip

{P r o o f \, o f \, f o r m u l a \, (\ref{der1}).} For the
function $f(\rho) = J(t_{R},\rho)- A\rho/4$ from
(\ref{th2.c}) we have
$$
f' = \frac{4t_{R}(1+t_{R})}
{\rho+ \sqrt{(1+2t_{R})^2\rho^2 - 4t_{R}(1+t_{R})}}- \frac{A}{4}\,,
\qquad f''(t,\rho) < 0\,.
$$
Then for $\rho \geq \tau_{R}$ we have
$$
f' \leq f'\Big|_{\rho = \tau_{R}} =
\frac{4t_{R}(1+t_{R})} {\tau_{R}}-\frac{A}{4} = \frac{\tau_{R}}
{1-\tau_{R}^{2}} -\frac{A}{4} \leq 0\,,
$$
if $\tau_{R} \leq \overline{\tau}_{1}(A)$, which proves the formula
(\ref{der1}). $\qquad \blacktriangle$

\medskip

{P r o o f \, o f \, l e m m a \, 1.} Let
$\{\mbox{\boldmath $x$}_{1}, \ldots, \mbox{\boldmath $x$}_{M}\}
\subset S^{n-1}(\sqrt{An})$ be a code such that
$\max\limits_{i \neq j}(\mbox{\boldmath $x$}_{i},
\mbox{\boldmath $x$}_{j}) \leq 0$, i.e. $\min\limits_{i \neq j}
\|\mbox{\boldmath $x$}_{i} -\mbox{\boldmath $x$}_{j}\|^{2} \geq 2A$.
Then, clearly, $M \leq 2n$.

In lemma 1 for all $i$ we have $\left(\mbox{\boldmath $x$}_{i},
\mbox{\boldmath $y$}\right) = (A+r-s)n/2$. Consider $M$ vectors
$\{\mbox{\boldmath $x$}_{i}' = \mbox{\boldmath $x$}_{i} -
a\mbox{\boldmath $y$}\}$, where $a = (A+r-s)/(2r)$. Then due to the
condition (\ref{condX1}) we have
$$
\max_{i \neq j}\left(\mbox{\boldmath $x$}_{i}',
\mbox{\boldmath $x$}_{j}'\right)
\leq \left[4Ar\rho - (A+r-s)^{2}\right]n/(4r) \leq 0\,,
$$
and therefore $M \leq 2n$. $\qquad \blacktriangle$

\medskip

{P r o o f \, o f \, l e m m a \, 2.} To prove lemma we reduce
it to the case $\rho \approx 0$, and then use lemma 4 (see
below). We set some integer $m$ such that $1 < m < M$, and introduce
the vector
$$
\mbox{\boldmath $z$} = a\sum_{k=1}^{m}\mbox{\boldmath $x$}_{k}\,,
\qquad a = \frac{\rho}{1+(m-1)\rho}\,.
$$
After simple calculations we get
\begin{equation}\label{ap1}
\begin{gathered}
\rho - \delta - \frac{1}{m} \leq
\left\|\mbox{\boldmath $z$}\right\|^{2} \leq \rho + \delta\,, \\
\frac{\rho - \delta}{1+(1-\rho)/(m\rho)} \leq
(\mbox{\boldmath $x$}_{i},\mbox{\boldmath $z$}) \leq
\frac{\rho + \delta}{1+(1-\rho)/(m\rho)} \,, \qquad i=m+1,\ldots,M.
\end{gathered}
\end{equation}
Consider the normalized vectors
$$
\mbox{\boldmath $u$}_{i} = \frac{\mbox{\boldmath $x$}_{i} -
\mbox{\boldmath $z$}}{\left\|\mbox{\boldmath $x$}_{i} -
\mbox{\boldmath $z$}\right\|}\,, \qquad i=m+1,\ldots,M.
$$
Using the formulas (\ref{ap1}), for any $i,j \geq m+1,\,i \neq j$,
we get
\begin{equation}\label{ap2}
(\mbox{\boldmath $u$}_{i},\mbox{\boldmath $u$}_{j}) \leq
\frac{2}{(1-\rho)}\left(\delta + \frac{1}{m}\right) = o(1)\,, \quad
n \to \infty\,,
\end{equation}
if we set $m \to \infty$ as $n \to \infty$. To upperbound the
maximal possible number $M-m$ of vectors
$\{\mbox{\boldmath $u$}_{i}\}$ satisfying the condition
(\ref{ap2}), we use a modification of \cite[Theorem 2]{Ran1}.

\medskip

{L e m m a \,4}. {\it Let ${\cal C} = \{\mbox{\boldmath $x$}_{1},
\ldots,\mbox{\boldmath $x$}_{M}\} \subset S^{n-1}(1)$ be a code with
$(\mbox{\boldmath $x$}_{i}, \mbox{\boldmath $x$}_{j}) \leq \mu,\,
i \neq j$. Then for $n \geq 1$ the upper bound holds}
\begin{equation}\label{Ran3}
\begin{gathered}
M \leq 2n^{3/2}(1-\mu)^{-n/2}\,, \qquad 0 \leq \mu < 1\,.
\end{gathered}
\end{equation}

\medskip

{P r o o f}. Denote $\mu = \cos (2\varphi)$, and let $M(\varphi)$ be
the maximal cardinality of such a code. For $M(\varphi)$ the upper
bound holds \cite[Theorem 2]{Ran1}
\begin{equation}\label{Ran1}
\begin{gathered}
M(\varphi) \leq \frac{(n-1)\sqrt{\pi}\,\Gamma\left(\dfrac{n-1}{2}
\right) \sin \beta \tan \beta}{2\Gamma\left(\dfrac{n}{2}\right)
\left[\sin^{n-1}\beta - f(\beta,n-2)\cos \beta\right]}\,,
\qquad 0 < \varphi < \frac{\pi}{4}\,,
\end{gathered}
\end{equation}
where $\beta = \arcsin(\sqrt{2}\sin \varphi)$ and
$$
f(\beta,n-2) = (n-1)\int\limits_{0}^{\beta}\sin^{n-2}z\,dz\,.
$$
Integrating by parts, for the function $f(\beta,n-2)$ we have
$$
\begin{gathered}
f(\beta,n-2) = \frac{\sin^{n-1}\beta}{\cos \beta} -
\frac{\sin^{n+1}\beta}{(n+1)\cos^{3}\beta} -
\frac{3}{(n+1)}\int\limits_{0}^{\beta}
\frac{\sin^{n+2}z}{\cos^{4}z}\,dz \geq \\
\geq \frac{\sin^{n-1}\beta}{\cos \beta} -
\frac{\sin^{n+1}\beta}{(n+1)\cos^{3}\beta} -
\frac{3\tan^{4}\beta}{(n+1)}f(\beta,n-2)\,,
\end{gathered}
$$
and therefore
\begin{equation}\label{evalf}
1\Big/\left[1+\dfrac{3\tan^{4}\beta}{n^{2}-1}\right] \leq
f(\beta,n-2)\Big/\left\{\frac{\sin^{n-1}\beta}{\cos \beta}
\left[1-\frac{\tan^{2}\beta}{n+1}\right]\right\} \leq 1\,,
\end{equation}
if $\tan^{2}\beta < n+1$, i.e. if $2\sin^{2}\varphi < (n+1)/(n+2)$.
From (\ref{Ran1}) and (\ref{evalf}) we get
\begin{equation}\label{evalf1}
\begin{gathered}
M(\varphi) \leq \frac{\sqrt{\pi}\,\Gamma\left(\dfrac{n-1}{2}\right)
(n^{2}-1)\cos\beta}{2\Gamma\left(\dfrac{n}{2}\right)\sin^{n-1}\beta}
< \frac{n\sqrt{\pi n(1-2\sin^{2}\varphi)}}
{\sqrt{2}\left(\sqrt{2}\sin \varphi\right)^{n-1}}\,,
\end{gathered}
\end{equation}
since
$$
\Gamma\left(\dfrac{z-1}{2}\right)(z^{2}-1)\bigg/
\Gamma\left(\dfrac{z}{2}\right) < \sqrt{2}\,z^{3/2}e^{1/z}\,,
\qquad z \geq 0\,.
$$
From (\ref{evalf1}) the inequality (\ref{Ran3}) follows provided
$2\sin^{2}\varphi < (n+1)/(n+2)$, i.e. if $\mu > 1/(n+2)$. Since the
function $M(\varphi)$ is continuous on the left for
$\varphi \in (0,\pi]$, the upper bound (\ref{evalf1}) remains valid
for $\mu = 1/(n+2)$ as well. For $\mu = 1/(n+2),\,n \geq 1$, the
right-hand side of (\ref{evalf1}) does not exceed $n\sqrt{\pi e/2}$,
which in turn does not exceed the right-hand side of (\ref{Ran3})
for any $\mu \geq 0,\,n \geq 2$. Since $M(\varphi)$ is a decreasing
function, it proves the inequality (\ref{Ran3}) for any
$\mu \geq 0,\,n \geq 2$. Clearly, (\ref{Ran3}) remains valid for
$n =1$ as well. $\qquad \blacktriangle$

Now from (\ref{ap2}) and (\ref{Ran3}) we get lemma 2.
$\qquad \blacktriangle$

\medskip

The author thanks L.A.Bassalygo, G.A.Kabatyansky and V.V.Prelov
for useful \\ discussions and constructive critical remarks.

\vspace{5mm}

\begin{flushleft}
{\small {\it Burnashev Marat Valievich} \\
Institute for Information Transmission Problems RAS \\
 {\tt burn@iitp.ru}}
\end{flushleft}%

\newpage

\includegraphics[width=0.9\hsize,height=0.8\hsize]{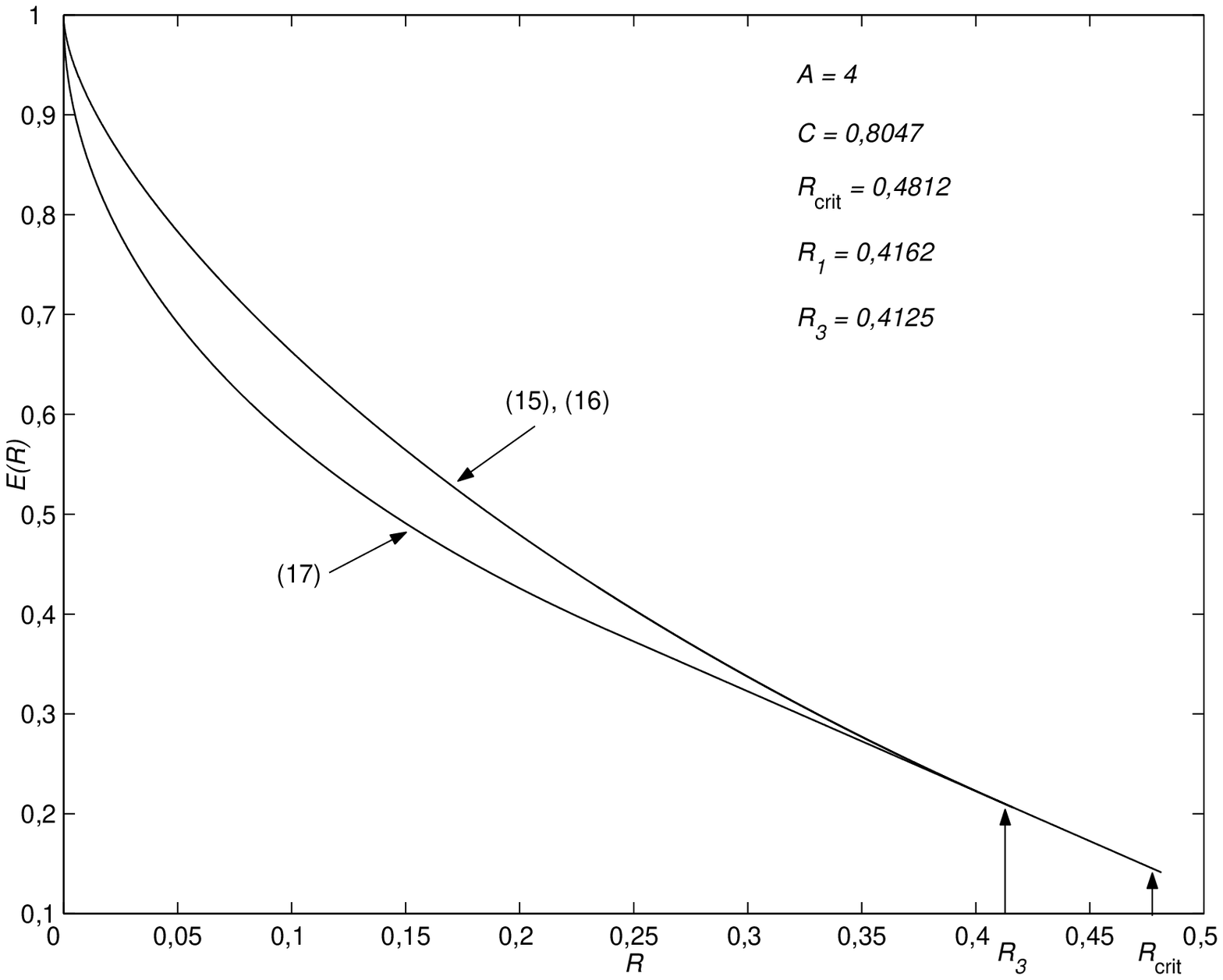}

%\smallskip
%\medskip

\begin{center} { Figure. Upper (15),(16) and lower (17) bounds
for $E(R,A)$ and $A = 4$}
\end{center}

\end{document}